\documentclass[10pt,aps,prresearch,twocolumn,superscriptaddress,floatfix]{revtex4-2}
\usepackage{amsmath}
\usepackage{amssymb}
\usepackage{mathrsfs}
\usepackage{bm}

\usepackage{graphicx}
\usepackage{dcolumn}

\usepackage{physics}

\usepackage{tikz}
\usetikzlibrary{calc,arrows,backgrounds}

\usepackage[colorlinks=true,allcolors=blue]{hyperref}

\usepackage{xcolor}

\begin{document}

\title{Spectral Signatures of Active Fluctuations in Semiflexible Polymers} 

\author{Love Grover}
\email{ph17047@iisermohali.ac.in}
\affiliation{Department of Physical Sciences, Indian Institute of Science Education and Research Mohali, Sector 81, Knowledge City, S. A. S. Nagar Manauli PO 140306, India}
\author{Anil Kumar Dasanna}
\email{adasanna@iisermohali.ac.in}
\affiliation{Department of Physical Sciences, Indian Institute of Science Education and Research Mohali, Sector 81, Knowledge City, S. A. S. Nagar Manauli PO 140306, India}
\author{Abhishek Chaudhuri}
\email{abhishek@iisermohali.ac.in}
\affiliation{Department of Physical Sciences, Indian Institute of Science Education and Research Mohali, Sector 81, Knowledge City, S. A. S. Nagar Manauli PO 140306, India}

\date{\today}

\begin{abstract}
We study how an active bath is transduced into the internal fluctuation spectrum of a semiflexible polymer. Starting from the statistics of active forces exerted by an explicit bath of active Brownian particles, we derive an effective description in terms of temporally persistent and spatially correlated noise, and test it against simulations of both explicit-bath and implicit-noise models. We find that activity reorganizes polymer fluctuations spectrally rather than uniformly: increasing the active force predominantly enhances the lowest modes, while increasing persistence shifts the spectral weight toward progressively longer wavelengths. The theory captures this mode-level reorganization well and explains the strong qualitative correspondence between explicit and implicit active baths over a broad parameter range. In contrast, global size measures such as the radius of gyration are systematically underestimated, which we trace to activity-induced bond stretching and contour-length renormalization absent from the present fixed-contour theory. Our results show that a semiflexible polymer acts as a multiscale probe of active matter, resolving the temporal and spatial structure of nonequilibrium forcing through its mode spectrum.
\end{abstract}

\maketitle


\section{Introduction}

The concept of temperature lies at the heart of equilibrium statistical mechanics,
providing a single parameter that characterizes fluctuations across all degrees of freedom.
In active matter---systems driven out of equilibrium by agents such as molecular motors,
ion pumps, or self-propelled particles---this universality generally breaks down.
Active systems are continuously fueled at the microscopic scale and they violate the
fluctuation--dissipation theorem (FDT) that underpins equilibrium thermodynamics~\cite{gnesotto2018broken,o2022time}.

A long-standing goal in the field has therefore been to understand to what extent
thermodynamic ideas can be extended to active systems.
Early experiments showed that passive tracers in
bacterial suspensions exhibit enhanced diffusion reminiscent of motion at elevated temperature~\cite{wu2000particle,gregoire2001comment,mino2011enhanced,chen2007fluctuations,angelani2011effective}. This motivated the introduction of an ``effective temperature'' ($T_\text{eff}$), often defined operationally through a ratio of fluctuation to dissipation~\cite{cugliandolo2011effective,loi2011effective}. For simple active particles and tracers, generalized equipartition-like relations can indeed yield useful effective temperatures~\cite{maggi2014generalized}. Recent studies have further characterized active Brownian particle systems in terms of
activity-dependent kinetic temperatures, effective diffusivities, and viscous drag, highlighting
both the utility and the limitations of reduced thermodynamic and transport descriptors in active
matter~\cite{palacci2010sedimentation,jepson2013enhanced,dal2019linear,cugliandolo2019effective,petrelli2020effective,hecht2024define,mandal2019motility,reichert2021tracer,burkholder2020nonlinear,caprini2020active,reichert2021tracer,khali2024active}.
However, such effective temperatures are typically observable- and timescale-dependent, so different probes need not report the same value. Moreover, even the active bath itself need not admit a unique thermal characterization: for example, in motility-induced phase separation (MIPS), dense and dilute phases can exhibit distinct effective temperatures~\cite{mandal2019motility}. An active environment, therefore, cannot, in general, be viewed as a single thermal reservoir~\cite{fodor2016far}.

This issue becomes even sharper for extended soft objects, whose internal degrees of freedom span
a hierarchy of relaxation times and length scales. Active fluctuations typically possess intrinsic temporal persistence~\cite{fodor2015activity,fodor2018statistical}, and may also be correlated spatially. As a result, the same active bath can act very differently on different internal modes of an embedded object. A semiflexible polymer is a minimal model system in which such scale-dependent coupling can be studied systematically. Its conformations can be decomposed into normal modes, each with its own relaxation time, so that the polymer naturally acts as a multiscale probe of active forcing. 

One common way to model an active environment is to replace it with temporally correlated noise~\cite{maes2020fluctuating}. A variety of colored-noise descriptions have been used in this context, including telegraphic or dichotomous noise, run-and-tumble forcing, and Ornstein--Uhlenbeck (OU) noise, both for active networks and for passive polymers driven out of equilibrium by active fluctuations
~\cite{gladrow2016broken,vandebroek2015active,samanta2016chain,baskaran2008enhanced,ghosh2014dynamics}.
Related analytical studies of flexible and semiflexible active polymers have shown that activity can strongly modify both conformational statistics and internal dynamics, with finite extensibility and mode-dependent relaxation playing an important role
~\cite{eisenstecken2016conformational,eisenstecken2017internal,eisenstecken2022path}.
More recently, colored-noise descriptions with explicit spatial structure have also been introduced to mimic forcing with a finite correlation length, as expected in active fluids and chaotic nonequilibrium media~\cite{frechette2025active}.

Complementary to such implicit descriptions, passive polymers in active baths have also been studied in a variety of explicit settings, including suspensions of self-propelled particles, hydrodynamic active environments, and motor-driven substrates, where activity has been shown to induce strong conformational changes, anomalous transport, localization, migration, and swelling or collapse
phenomena~\cite{ananthakrishnan2007forces,winkler2020physics,harder2014role,nikola2016active,winkler2017active,shin2015facilitation,anderson2022polymer,aporvari2020anisotropic,mousavi2021active,muzzeddu2024migration,weady2024conformations,paul2022activity,martin2020hydrodynamics,foglino2019non,gupta2019morphological,shee2021semiflexible}.
These studies collectively show that active forcing can often be represented at a coarse-grained level by a temporally persistent and spatially structured drive, but they also highlight the wide range of conformational responses that can arise in explicit nonequilibrium environments. At the same time, an important question remains open: which features of the response of a polymer in an \emph{explicit} active bath can be captured by a reduced implicit colored-noise description, and which require additional geometric or microscopic information beyond such a coarse-grained representation?

This question is most naturally addressed through a mode-resolved analysis.
Historically, shape-mode decompositions have been widely used to extract mechanical properties of semiflexible filaments. For example, Gittes et al.~\cite{gittes1993flexural} and Brangwynne et al.~\cite{brangwynne2007bending} used mode amplitudes to infer the flexural rigidity of microtubules and actin by relating thermal mode fluctuations to equipartition. In active systems, however, mode analysis reveals much richer physics.
Brangwynne et al.~\cite{brangwynne2008nonequilibrium} showed that intracellular activity strongly amplifies microtubule shape fluctuations beyond equilibrium expectations, while Gladrow et al.~\cite{gladrow2016broken} demonstrated that active filament networks violate detailed balance and exhibit non-trivial energy circulation in mode space.
These studies show that mode amplitudes can be used to resolve the distribution of nonequilibrium forcing across scales.

\begin{figure}[t]
    \centering
    \begin{tikzpicture}[scale=1.0]
    \tikzstyle{bead} = [circle, draw=black, fill=blue!20, minimum size=1cm, inner sep=0pt, outer sep=0pt]
    
    \tikzstyle{activeSmall} = [circle, draw=red!80!black, fill=red!30, minimum size=0.3cm, inner sep=0pt]
    \tikzstyle{activeBig} = [circle, draw=red!80!black, fill=red!30, minimum size=0.6cm, inner sep=0pt]
    
    \tikzstyle{arrow} = [->, >=latex, thick, red!80!black]

    
    \node[bead] (b1) at (0,0) {};
    \node[bead] (b2) at (10:1.0) {};
    \path (b2) ++(-15:1.0) node[bead] (b3) {};
    \path (b3) ++(-30:1.0) node[bead] (b4) {};
    \path (b4) ++(-10:1.0) node[bead] (b5) {};
    \path (b5) ++(10:1.0) node[bead] (b6) {};
    \path (b6) ++(30:1.0) node[bead] (b7) {};


    
    \node[activeBig] (a1) at (-0.8, 1.2) {};
    \draw[arrow] (a1) -- ++(-45:0.8);
    
    \node[activeSmall] (a2) at (1.2, 0.8) {};
    \draw[arrow] (a2) -- ++(-90:1.2);

    \node[activeBig] (a3) at (2.5, 0.8) {};
    \draw[arrow] (a3) -- ++(180:0.4);

    \node[activeSmall] (a4) at (1.0, -1.2) {\tiny $i$};
    \draw[arrow] (a4) -- ++(60:0.6) node[midway, right] {$f_a^{i}$};

    \node[activeBig] (a5) at (3.0, -1.8) {};
    \draw[arrow] (a5) -- ++(120:0.5);

    \node[activeSmall] (a6) at (5.0, 1.0) {};
    \draw[arrow] (a6) -- ++(-135:0.9);

    \node[activeBig] (a7) at (6.5, 0.5) {};
    \draw[arrow] (a7) -- ++(160:0.7);

    \node[activeSmall] (a8) at (-1.5, -0.5) {};
    \draw[arrow] (a8) -- ++(30:0.5);

    \node[activeSmall] (a9) at (6.0, -1.5) {};
    \draw[arrow] (a9) -- ++(100:0.6);

    \node[activeBig] (a10) at (3.5, 2.0) {};
    \draw[arrow] (a10) -- ++(-60:0.4);

    \node[activeBig] (a11) at (-0.2, -1.8) {};
    \draw[arrow] (a11) -- ++(45:0.7);
    
    \node[activeSmall] (a12) at (4.0, 0.8) {\tiny $j$};
    \draw[arrow] (a12) -- ++(-110:0.8) node[midway, right] {$f_a^{j}$};

    \node[activeSmall] (a13) at (0.5, 2.2) {};
    \draw[arrow] (a13) -- ++(-20:0.5);

    \node[activeBig] (a14) at (5.55, -0.9) {};
    \draw[arrow] (a14) -- ++(150:0.6);
    
    \node[activeSmall] (a15) at (-1.0, 0.5) {};
    \draw[arrow] (a15) -- ++(10:0.5);

\end{tikzpicture}
    \caption{Schematic representation of a semiflexible polymer (blue beads) in a busy active environment.
    The polymer consists of beads of diameter $\sigma$.
    The active particles (red circles) are self-propelled with persistence vectors (arrows).
    Their collisions and persistent pushing generate forcing with nontrivial time and length scales on the polymer. While this schematic illustrates a generic active medium, our study specifically employs a monodisperse active bath with uniform self-propulsion.
    }
    \label{fig:polymer_concept}
\end{figure}
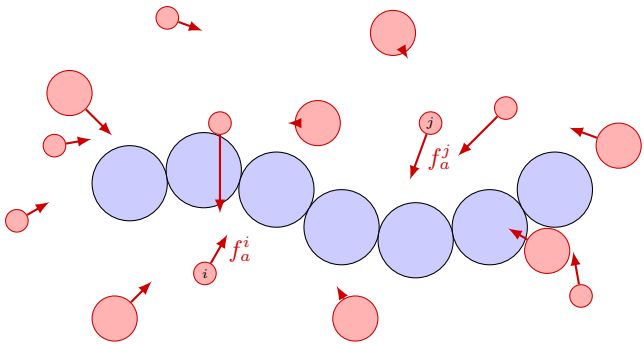

In this work, we study a semiflexible polymer immersed in an explicit bath of active Brownian particles and ask how the bath affects the polymer’s internal fluctuation spectrum. Starting from the statistics of the active forces acting on the polymer, we construct an approximate effective description in which the bath acts through temporally persistent, spatially correlated forcing. We then test this description primarily against simulations of a polymer embedded in an explicit active-particle bath. In addition, we compare these results with a reduced implicit coloured-noise model.

Our central result is that the active bath reorganizes polymer fluctuations spectrally rather than uniformly. At fixed persistence time, increasing the active force magnitude $f_a$ selectively enhances the low-mode sector, while at fixed $f_a$, increasing the persistence time $\tau$ shifts the spectral weight toward progressively lower mode number. The active bath, therefore, does not act as a uniform heat source. It instead couples
preferentially to modes whose intrinsic relaxation times are comparable to or longer than the bath correlation time. We show that the reduced theory captures this mode-level reorganization well for the explicit-bath model. We further find that the implicit coloured-noise model exhibits qualitatively similar large-scale conformational trends across a broad region of parameter space. We do not attempt a strict one-to-one mapping of microscopic parameters between the two descriptions.

Our analysis naturally leads to a mode-dependent thermodynamic description. By mapping the active enhancement of each mode onto the equilibrium mode spectrum of a passive
polymer, one may assign mode-dependent effective temperatures rather than a single scalar temperature. From this perspective, the polymer acts as a spectroscopic probe of the active environment~\cite{singh2024anomalous}, resolving how different wavelengths couple differently to nonequilibrium forcing. More broadly, our results show that the relevant time and length scales of active forcing are filtered through the polymer’s hierarchy of internal modes, and that this filtering determines its nonequilibrium conformational state.

\section{Models and simulation protocol}
\label{sec:model}

We study a semiflexible polymer driven out of equilibrium by two distinct realizations of activity: (i) an \emph{explicit} bath of self-propelled particles (active Brownian particles, ABPs) that exerts forces through direct interactions, and (ii) an \emph{implicit} bath in which activity enters as a prescribed Ornstein--Uhlenbeck (OU) colored-noise forcing acting on the polymer beads. Unless otherwise stated, all simulations are performed in two dimensions with periodic boundary conditions.

\subsection{Polymer model and conservative interactions}
\label{sec:polymer_model}

The polymer is represented as a bead--spring chain of $N$ beads (diameter $\sigma$) at positions $\{\mathbf{r}_i\}_{i=1}^N$.
Nearest-neighbour connectivity is enforced by harmonic bonds,
\begin{equation}
\label{bonds}
V_{b}=\frac{k_s}{2}\sum_{i=2}^{N}\left(\,|\mathbf{r}_{i}-\mathbf{r}_{i-1}|-r_0\,\right)^2,
\end{equation}
and semiflexibility is introduced via a harmonic bending penalty on successive bond angles,
\begin{equation}
\label{bending}
V_{a}={\kappa}\sum_{i=2}^{N-1}\left(1-\cos{\theta_i}\right),
\end{equation}
where $\theta_i$ is the angle between bond vectors $\mathbf{r}_{i}-\mathbf{r}_{i-1}$ and $\mathbf{r}_{i+1}-\mathbf{r}_{i}$. 
The equilibrium bond length is $r_0=2^{1/6}\sigma$.

Excluded-volume interactions are modeled by a truncated (purely repulsive) Weeks--Chandler--Andersen (WCA) potential. For any interacting pair at separation $r$,
\begin{equation}
\label{LJ_dimless}
V_{\rm LJ}(r)=
4\epsilon\left[\left(\frac{\sigma}{r}\right)^{12}-\left(\frac{\sigma}{r}\right)^{6}\right],
\qquad r<r_{\rm cut},
\end{equation}
with $r_{\rm cut}=2^{1/6}\sigma$. In the explicit-bath model, the same WCA form is used for polymer-polymer (non-bonded), polymer-ABP, and ABP-ABP interactions. We denote by $\mathbf{F}^{\rm c}_i\equiv-\nabla_{\mathbf{r}_i}(V_b+V_a+V_{\rm WCA})$ the total conservative force on polymer bead $i$.

\subsection{Explicit bath model (ABP bath)}
\label{sec:explicit_bath}

In the explicit-bath model, the polymer is immersed in a bath of $N_a$ active Brownian particles (disks of diameter $\sigma$) at number density
$\rho_0=N_a/A$ in a simulation box of area $A$.
Each ABP $j$ has position $\mathbf{R}_j$ and orientation $\hat{\mathbf{n}}_j=(\cos\theta_j,\sin\theta_j)$.

\textit{Polymer dynamics in the explicit bath.}
Polymer beads evolve via the underdamped Langevin equation,
\begin{equation}
\label{langevinpolymer}
m\frac{d\mathbf{v}_i}{dt}=-\gamma_t\,\mathbf{v}_i+\mathbf{F}^{\rm c}_i+\sqrt{2\gamma_t k_B T}\,\boldsymbol{\xi}_i(t),
\end{equation}
where $\mathbf{v}_i=d\mathbf{r}_i/dt$, $\gamma_t$ is the translational friction coefficient, and
$\boldsymbol{\xi}_i(t)$ is Gaussian white noise with
$\langle \xi_{i\alpha}(t)\rangle=0$ and $\langle \xi_{i\alpha}(t)\xi_{j\beta}(t')\rangle=\delta_{ij}\delta_{\alpha\beta}\delta(t-t')$.

\textit{Active-particle dynamics.}
Each ABP experiences conservative forces from WCA interactions with other ABPs and with polymer beads, plus a constant-magnitude self-propulsion force:
\begin{equation}
\label{langevinactive}
m_a\frac{d\mathbf{V}_j}{dt}=-\gamma_t\,\mathbf{V}_j+\mathbf{F}^{\rm c}_j+f_a\,\hat{\mathbf{n}}_j+\sqrt{2\gamma_t k_B T}\,\boldsymbol{\Xi}_j(t),
\end{equation}
where $\mathbf{V}_j=d\mathbf{R}_j/dt$, $\mathbf{F}^{\rm c}_j$ is the total conservative force on ABP $j$, and $\boldsymbol{\Xi}_j(t)$ is Gaussian white noise
(with correlations analogous to $\boldsymbol{\xi}_i(t)$).


The particle orientation is characterized, at the coarse-grained level, by rotational diffusion,
\begin{equation}
\label{langevinactiveangle}
\frac{d\theta_j}{dt}=\sqrt{2D_r}\,\eta_j(t),
\end{equation}
where $\langle \eta_j(t)\rangle=0$ and
$\langle \eta_j(t)\eta_{j'}(t')\rangle=\delta_{jj'}\delta(t-t')$.
The corresponding persistence time of active motion is $\tau= {1}/{D_r}.$ In the simulations, this orientational dynamics is implemented using an underdamped rotational Langevin thermostat~\cite{dias2021molecular} rather than the overdamped form of
Eq.~\eqref{langevinactiveangle}. We use Eq.~\eqref{langevinactiveangle} only as the effective coarse-grained description that defines the persistence time $\tau$. The mapping between the simulation parameter $\alpha\equiv\textit{angmom}$ and the corresponding effective persistence time is given in Appendix~\ref{app:angmom}.

\subsection{Implicit bath model (OU colored forcing)}
\label{sec:implicit_bath}

As a reduced reference description, we also consider an implicit active-bath model in which the
polymer is driven by temporally correlated noise rather than by explicit active particles.
In this model, there are no surrounding ABPs. Instead, each polymer bead experiences an additional
Ornstein--Uhlenbeck (OU) force $\mathbf{F}^{\rm OU}_i(t)$, taken to be independent across beads.
The bead dynamics is then
\begin{equation}
\label{langevinou}
m\frac{d\mathbf{v}_i}{dt}
=
-\gamma_t\,\mathbf{v}_i
+\mathbf{F}^{\rm c}_i
+\mathbf{F}^{\rm OU}_i(t)
+\sqrt{2\gamma_t k_B T}\,\boldsymbol{\xi}_i(t),
\end{equation}
where $\mathbf{F}^{\rm c}_i$ denotes the conservative bead--spring and bending forces, and
$\boldsymbol{\xi}_i(t)$ is Gaussian white thermal noise.

The OU forcing is specified by zero mean and exponentially decaying temporal correlations,
\begin{equation}
\label{ouforce}
\big\langle \mathbf{F}^{\rm OU}_i(t)\cdot \mathbf{F}^{\rm OU}_j(t')\big\rangle
=
(f_a^{\rm OU})^2\,e^{-|t-t'|/\tau^{\rm OU}}\,\delta_{ij},
\end{equation}
where $f_a^{\rm OU}$ sets the forcing amplitude and $\tau^{\rm OU}$ is the persistence time of
the colored noise. Equivalently, the OU force may be generated from
\[
\dot{\mathbf{F}}^{\rm OU}_i
=
-\frac{1}{\tau^{\rm OU}}\mathbf{F}^{\rm OU}_i
+
\sqrt{\frac{2(f_a^{\rm OU})^2}{\tau^{\rm OU}}}\,\boldsymbol{\eta}_i(t),
\]
with Gaussian white noise $\boldsymbol{\eta}_i(t)$.

In discrete time, the OU process can be advanced exactly over a timestep $\Delta t$ \cite{gillespie1996exact}:
\begin{equation}
\label{eq:ou_update}
\mathbf{F}_i^{\rm OU}(t+\Delta t)
=
e^{-\Delta t/\tau^{\rm OU}}\mathbf{F}_i^{\rm OU}(t)
+
f_a^{\rm OU}\sqrt{1-e^{-2\Delta t/\tau^{\rm OU}}}\;\mathbf{R}_i,
\end{equation}
where $\mathbf{R}_i$ is a vector of independent standard normal variates, drawn independently for
each bead and timestep.

The purpose of this model is not to reproduce the explicit active bath at a microscopic level, but
to provide a minimal temporally persistent forcing against which the explicit-bath results may be
qualitatively compared. In the present work, the analytical theory is fitted only to the
explicit-bath model. The implicit OU model is used solely as a reduced reference description to
assess whether a simple colored-noise forcing captures the broad conformational trends of the
explicit bath, rather than to establish a strict one-to-one parameter mapping. A qualitative comparison of the conformational response in the explicit and implicit models, based on the normalized radius-of-gyration landscape in the $(f_a,\tau)$ plane, is shown in Appendix~\ref{app:implicit_compare}.

\subsection{Simulation protocol and measured observables}
\label{sec:sim_protocol}


All simulations are performed in two dimensions with periodic boundary conditions using the LAMMPS. 
All physical quantities are expressed in reduced Lennard-Jones (LJ) units, where we set the fundamental units of length, energy, and mass to $\sigma = 1$, $k_B T = 1$, and $m_0 = 1$, respectively. This naturally defines the intrinsic unit of time as $\tau_{\rm LJ} = \sigma\sqrt{m_0/k_B T}$. 
Equations~\eqref{langevinpolymer}--\eqref{langevinactiveangle} for the explicit bath and Eq.~\eqref{langevinou}, together with the exact OU update~\eqref{eq:ou_update}, for the implicit bath are integrated numerically using a velocity-Verlet-based underdamped Langevin thermostat with a timestep $\Delta t = 0.002\,\tau_{\rm LJ}$. The simulation parameters are summarized in Table~\ref{tab:params}.

The polymer consists of $N = 64$ beads of diameter $\sigma = 1$ and mass $m = 2/3$, connected by harmonic bonds with spring constant $k_s / k_B T = 1000\,\sigma^{-2}$ and equilibrium length $r_0 = 2^{1/6}\sigma$. Bending stiffness is introduced via a cosine angle potential $V_a = \kappa(1 + \cos\theta)$, which reduces in the small-angle limit to the harmonic form $\kappa(\theta - \pi)^2 / 2$ used in the theory. We study two values of the bending stiffness: $\kappa = 32\,k_B T$ and $\kappa = 128\,k_B T$. All non-bonded interactions use a purely repulsive Weeks-Chandler-Andersen (WCA) potential with $\epsilon = 10\,k_B T$ and a cutoff radius $r_c = 2^{1/6}\sigma$.

Each bead evolves subject to a background friction $\gamma_t = 10$ and temperature $T = 1$, yielding an inertial relaxation time $m/\gamma_t \approx 0.067\,\tau_{\rm LJ}$. The explicit Active Brownian Particle (ABP) bath consists of particles of mass $m_a = 2/3$ at an area fraction $\phi = 0.1$, contained in a square box of side $L_{\rm box} = 128\,\sigma$. Self-propulsion is applied via a constant body-frame force of magnitude $f_a$, while the orientational persistence time $\tau = 1/D_r$ is regulated as described in Appendix~\ref{sec:explicit_bath}. We sweep the active force $f_a \in \{0, 8, 16, 32, 48, 64, 72, 80, 88, 96, 104\}$ and the persistence time $\tau \in \{0.1, 1, 4, 7, 10, 13, 16, 19\}$ (in LJ units).

Each run is initialized with a straight polymer in respective activity source and temperature set to $T$ in form of Langevin noise. Activity is introduced in the system either by introducing the ABP bath or by applying the OU forcing to each monomer. We simulate for a total of $10^8$ timesteps ($200,000\,\tau_{\rm LJ}$), with statistics accumulated in the resulting nonequilibrium steady state over a production time $t_{\rm prod} = 40,000\,\tau_{\rm LJ}$ (the last 20\% of the trajectory). Unless otherwise stated, all reported quantities are averaged over $N_{\rm run}$ independent realizations.

\begin{table}[h]
\centering
\caption{{Summary of simulation parameters (in LJ units, $k_BT=1$).}}
\label{tab:params}
\begin{tabular}{lll}
\hline\hline
{Parameter }& {Symbol }& {Value }\\
\hline
{Number of beads }& {$N$ }& {$64$ }\\
{Bead diameter }& {$\sigma$ }& {$1$ }\\
{Bead mass }& {$m$ }& {$2/3$ }\\
{Temperature }& {$T$ }& {$1$ }\\
{Friction coefficient }& {$\gamma_t$ }& {$10$ }\\
{Timestep }& {$\Delta t$ }& {$0.002$ }\\
{Bond spring constant }& {$k_s$ }& {$1000$ }\\
{Equilibrium bond length }& {$r_0$ }& {$2^{1/6}\sigma$ }\\
{WCA energy }& {$\epsilon$ }& {$10$ }\\
{Bending stiffness }& {$\kappa$ }& {$32,\ 128$ }\\
{Box size }& {$L_{\rm box}$ }& {$128\sigma$ }\\
{ABP mass }& {$m_a$ }& {$2/3$ }\\
{ABP area fraction }& {$\phi$ }& {$0.1$ }\\
{Activity strength }& {$f_a$ }& {$0$--$104$ }\\
{Persistence time }& {$\tau$ }& {$0.1$--$19$ }\\
{Number of realizations }& {$N_{\rm run}$ }& {$8$ }\\
{Run length }& & {$10^8$ steps }\\
\hline\hline
\end{tabular}
\end{table}

The main observables considered in this work are the polymer mode amplitudes and their steady-state
variances, from which we construct the mode-resolved effective temperatures introduced in
Sec.~\ref{sec:theory}. We also measure the radius of gyration,
\[
R_g^2=\frac{1}{N}\sum_{i=1}^{N}\left|\mathbf r_i-\mathbf r_{\rm cm}\right|^2,
\qquad
\mathbf r_{\rm cm}=\frac{1}{N}\sum_{i=1}^{N}\mathbf r_i.
\]
These quantities allow us to compare the spectral reorganization of polymer fluctuations with the
corresponding global conformational response.


Parameter sweeps are performed over $(f_a,\tau)$ for the explicit bath and over
$(f_a^{\rm OU},\tau^{\rm OU})$ for the implicit OU model. In the present work, the analytical
theory is compared quantitatively only with the explicit-bath simulations. The implicit model is
used solely as a reduced reference description to assess whether a minimal colored-noise forcing
captures the broad conformational trends of the explicit bath. Accordingly, no strict
parameter-by-parameter mapping between the two models is imposed here; comparisons between them are
restricted to qualitative trends in observables such as the mode spectrum and the activity-induced
variation of $R_g$. A corresponding qualitative comparison of $R_g$ over parameter space is provided in
Appendix~\ref{app:implicit_compare}.

\section{Theoretical Framework}
\label{sec:theory}

We now develop a microscopic route from an explicit bath of active Brownian particles (ABPs) to an
\emph{effective} colored-noise forcing acting on a passive semiflexible polymer.

\subsection{System parameters and ABP dynamics}
\label{subsec:abp_dynamics}

We consider a passive semiflexible polymer of contour length $L$ (discretized into $N$ beads in the simulations), bending rigidity $\kappa$, and local friction coefficient $\zeta\approx\gamma_t$ in two spatial dimensions ($d=2$). The polymer is immersed in a bath of active Brownian particles (ABPs) at number density $\rho_0$. In the simulations, this corresponds to an area fraction $\phi=\rho_0\pi\sigma^2/4=0.1$, or $\rho_0\approx0.127\sigma^{-2}$. Each ABP is self-propelled with speed $v_a=f_a/\gamma_t$, where $f_a$ is the active force magnitude (corresponding to the simulation parameter $f_a$), and $\gamma_t$ is the particle friction, and undergoes rotational diffusion with rate $D_r$. The corresponding orientational persistence time is $\tau \equiv {1}/{D_r}$.
Polymer--particle interactions are short-ranged and repulsive, described by a pair potential $V(r)$ of characteristic range $\sigma$, with associated force field $\mathbf f(\mathbf x)\equiv -\nabla V(|\mathbf x|)$.

In our analytical approach, we begin from the overdamped free-space dynamics of an ABP for simplicity:
\begin{align}
\dot{\mathbf R}(t) &= v_a \hat{\mathbf u}(t) + \sqrt{2D_t}\,\boldsymbol\xi(t), \label{eq:abp_r_free_revised}\\
\dot{\theta}(t) &= \sqrt{2D_r}\,\eta(t),
\qquad
\hat{\mathbf u}(t)=(\cos\theta,\sin\theta), \label{eq:abp_theta_revised}
\end{align}
with independent Gaussian white noises $\boldsymbol\xi(t)$ and $\eta(t)$ of unit variance.
Here $D_t=\mu k_B T$ is the translational diffusivity of an isolated ABP.

In the explicit-bath simulations, ABPs interact with polymer beads through short-ranged
excluded-volume forces, so their motion is not strictly free over all times. However, for the
purpose of deriving the \emph{form} of the effective active forcing, we use the free-space ABP statistics as the microscopic input to the Gaussian-displacement closure developed below. This approximation should therefore be understood as a reduced description of bath-particle motion over the interaction zone, rather than as a complete model of the explicit simulations. Its consequences and limitations are discussed in the following sections.

\subsection{Displacement statistics of a free ABP}
\label{subsec:msd}

The only ingredient from the free-ABP dynamics needed below is the displacement covariance over a
time lag $\Delta t$,
\[
\Delta \mathbf R(\Delta t)\equiv \mathbf R(t+\Delta t)-\mathbf R(t).
\]
For the overdamped ABP dynamics in Eqs.~\eqref{eq:abp_r_free_revised}--\eqref{eq:abp_theta_revised},
the displacement statistics are isotropic and satisfy
\[
\langle \Delta R_\alpha(\Delta t)\Delta R_\beta(\Delta t)\rangle
=
2S(\Delta t)\,\delta_{\alpha\beta},
\]
with
\begin{equation}
\label{eq:S_def_revised_short}
S(\Delta t)
=
D_t|\Delta t|
+
\frac{v_a^2\tau}{d}
\left(
|\Delta t|-\tau\left[1-e^{-|\Delta t|/\tau}\right]
\right).
\end{equation}
Equivalently,
\[
\langle |\Delta \mathbf R(\Delta t)|^2\rangle = 2d\,S(\Delta t).
\]

At short times, $S(\Delta t)\simeq D_t|\Delta t|+\frac{v_a^2}{2d}\Delta t^2$, while at long times
it crosses over to diffusive growth with effective diffusivity
\[
D_{\rm eff}=D_t+\frac{v_a^2\tau}{d}.
\]
In what follows, $S(\Delta t)$ serves as the microscopic input to the Gaussian-displacement closure
used to construct the active force correlator.

\subsection{Statistics of the effective active force on the polymer}
\label{subsec:force_statistics}

To characterize the forcing exerted by an explicit bath of active Brownian particles (ABPs) on the polymer, we begin from the microscopic interaction energy
\[
U=\sum_{i=1}^{N_a}\int_0^L ds\,V\!\left(|\mathbf R_i(t)-\mathbf r(s,t)|\right),
\]
which gives the active force density along the contour as
\begin{equation}
\label{eq:F_act_def_revised}
\mathbf F_{\rm act}(s,t)
=
-\frac{\delta U}{\delta \mathbf r(s,t)}
=
\sum_{i=1}^{N_a}\mathbf f\!\left(\mathbf R_i(t)-\mathbf r(s,t)\right),
\end{equation}
where $\mathbf f(\mathbf x)\equiv -\nabla V(|\mathbf x|)$.

In a homogeneous dilute bath, the mean force vanishes by symmetry, and the leading statistical information is contained in the connected force--force correlator
\[
C_{\alpha\beta}(s,t;s',t')
\equiv
\left\langle F_{{\rm act},\alpha}(s,t)\,F_{{\rm act},\beta}(s',t')\right\rangle_c .
\]
Neglecting bath--bath correlations beyond the uniform density $\rho_0$ and retaining only the single-particle contribution gives (see Appendix~\ref{app:forcecorr} for details):
\begin{multline}
\nonumber
C_{\alpha\beta}(s,t;s',t')
\simeq
\rho_0\int d^d\mathbf R \int d^d\mathbf R'\,
f_\alpha(\mathbf R-\mathbf r(s,t))\\
f_\beta(\mathbf R'-\mathbf r(s',t'))
P(\mathbf R',t'|\mathbf R,t),
\end{multline}
where $P(\mathbf R',t'|\mathbf R,t)$ is the single-particle propagator of a bath particle.

It is convenient to work in Fourier space. Using
\[
f_\alpha(\mathbf x)=\int \frac{d^d\mathbf k}{(2\pi)^d}\,\tilde f_\alpha(\mathbf k)\,
e^{i\mathbf k\cdot \mathbf x},
\]
and approximating the propagator by its free-space form
$P(\mathbf R',t'|\mathbf R,t)\approx P(\mathbf R'-\mathbf R,\Delta t)$ with
$\Delta t=t'-t$, one obtains after integrating over the center-of-mass coordinate
\begin{multline}
\label{eq:C_kspace_revised}
C_{\alpha\beta}(s,t;s',t')
=
\rho_0\int\frac{d^d\mathbf k}{(2\pi)^d}\,
\tilde f_\alpha(\mathbf k)\tilde f_\beta(-\mathbf k)\,
e^{i\mathbf k\cdot \Delta \mathbf r}\,\\
\left\langle e^{-i\mathbf k\cdot \Delta \mathbf R(\Delta t)}\right\rangle ,
\end{multline}
where $\Delta \mathbf r \equiv \mathbf r(s',t')-\mathbf r(s,t)$, and $\Delta \mathbf R(\Delta t)\equiv \mathbf R(t')-\mathbf R(t)$. The remaining average is the characteristic function of ABP displacements. We approximate it by
truncating the cumulant expansion at second order,
\[
\left\langle e^{-i\mathbf k\cdot \Delta \mathbf R(\Delta t)}\right\rangle
\simeq
\exp\!\left[-\frac12 k_\mu k_\nu
\left\langle \Delta R_\mu(\Delta t)\Delta R_\nu(\Delta t)\right\rangle\right].
\]
For an isotropic bath,
\[
\left\langle \Delta R_\mu(\Delta t)\Delta R_\nu(\Delta t)\right\rangle
=
2S(\Delta t)\delta_{\mu\nu},
\]
so that
\[
\left\langle e^{-i\mathbf k\cdot \Delta \mathbf R(\Delta t)}\right\rangle
\simeq e^{-k^2 S(\Delta t)}.
\]
Substituting this into Eq.~\eqref{eq:C_kspace_revised} gives
\[
C_{\alpha\beta}(\Delta \mathbf r,\Delta t)
=
\rho_0\int\frac{d^d\mathbf k}{(2\pi)^d}\,
\tilde f_\alpha(\mathbf k)\tilde f_\beta(-\mathbf k)\,
e^{i\mathbf k\cdot \Delta \mathbf r}\,
e^{-k^2 S(\Delta t)} .
\]

We assume that the interaction can be modeled as a Gaussian soft repulsion, $V(r)=V_0 e^{-r^2/(2\sigma^2)}$. Then the $k$-integral can be performed analytically. The resulting correlator is
\begin{multline}
\label{eq:C_realspace_revised}
C_{\alpha\beta}(\Delta \mathbf r,\Delta t)
=
\rho_0 V_0^2 \sigma^{2d}\left(\frac{\pi}{A(\Delta t)}\right)^{d/2}
\left[
\frac{\delta_{\alpha\beta}}{2A(\Delta t)}\right. \\
\left. -\frac{\Delta r_\alpha\Delta r_\beta}{4A(\Delta t)^2}
\right]
\exp\!\left[-\frac{|\Delta \mathbf r|^2}{4A(\Delta t)}\right],
\end{multline}
where $A(\Delta t)=\sigma^2+S(\Delta t)$. The derivation of Eq.~\eqref{eq:C_realspace_revised}, together with the explicit evaluation of
\(S(\Delta t)\) for ABPs, is given in Appendix~\ref{app:forcecorr}.

Equation~\eqref{eq:C_realspace_revised} already contains the essential qualitative features of the
explicit-bath forcing: it is short-ranged in space, has a finite memory inherited from ABP
displacements, and is non-separable in space and time through the combination
\(A(\Delta t)=\sigma^2+S(\Delta t)\). Rather than attempting to reduce
Eq.~\eqref{eq:C_realspace_revised} directly to a simple separable kernel in real space, we proceed
in a way more natural for the polymer problem and first project the forcing onto polymer modes.

\paragraph*{From the microscopic kernel to an effective mode-space forcing.}
In the weak-bending regime, the contour is nearly straight on the scales entering the force
correlator, so that
\[
|\mathbf r(s)-\mathbf r(s')|^2 \simeq (s-s')^2 .
\]
The correlator then becomes approximately translationally invariant along the contour,
\(C(s,t;s',t')\simeq C(s-s',\Delta t)\), and is therefore most naturally analyzed after projection
onto tangent modes. As shown in Appendix~\ref{app:mode_projection}, this projection leads
approximately to a mode-diagonal active forcing with a finite wave-number cutoff and a nontrivial
memory kernel. Guided by that projected structure, we adopt in the main text the following minimal
effective closure at the level of the \emph{projected} active force,
\begin{equation}
\label{eq:mode_kernel_main}
\left\langle \xi_n^{u,{\rm act}}(t)\,\xi_m^{u,{\rm act}}(t')\right\rangle
=
\delta_{nm}\,
q_n^2 D_0\,e^{-q_n^2\sigma^2}\,
e^{-|\Delta t|/\tau_{\rm eff}} .
\end{equation}
Here \(D_0\), \(\sigma\), and \(\tau_{\rm eff}\) are effective parameters. The factor
\(e^{-q_n^2\sigma^2}\) comes from the finite spatial range of the microscopic force kernel. 
The exponential memory is a compact single-timescale approximation to the exact temporal kernel. Thus, Eq.~\eqref{eq:mode_kernel_main} serves as an effective
\emph{mode-space} closure. This mode-space kernel is sufficient for the analytical theory developed below, and is used for the comparisons with numerical simulations.

\subsection{Tangent correlation spectrum}
\label{subsec:tangent_spectrum}

We now derive the mode-resolved fluctuation spectrum of the polymer. We shall compare the explicit simulations and the theory at the level of tangent fluctuations. Therefore, it is convenient to work with the transverse tangent field \(u_\perp(s,t)\) in the weak-bending regime. 

While the simulations employ full underdamped Langevin dynamics to capture the complete microscopic evolution of the system, we treat the analytical derivation of the mode spectrum in the overdamped limit. This approximation is physically justified for the large-scale conformational fluctuations of interest. Specifically, the intrinsic relaxation times of the structurally dominant long wavelength modes ($n = 1 \dots 32$) scale as $\tau_n= \zeta/(\kappa q_n^4)$, which are orders of magnitude larger than the inertial relaxation timescale $ m/\gamma_t\approx 0.067 \tau_{LJ}$ . On these relevant timescales, inertial effects effectively average out, allowing the overdamped theory to provide a consistent analytical approximation of the full simulated dynamics. The linearized overdamped dynamics take the form: 
\begin{equation}
\label{eq:u_eq_main}
\zeta\,\partial_t u_\perp(s,t)
=
-\kappa\,\partial_s^4 u_\perp(s,t)
+\eta_u^{\rm th}(s,t)
+\eta_u^{\rm act}(s,t),
\end{equation}
where \(\eta_u^{\rm th}=\partial_s\eta_{\rm th}\) and
\(\eta_u^{\rm act}=\partial_s F_{\rm act}\) denote the thermal and active tangent forcings.
At this level, nonlinear tension and inextensibility couplings are neglected; they are discussed
separately below.

For a free filament, we expand the tangent field in the cosine basis
\[
u_\perp(s,t)=\sum_{n\ge 1} a_n(t)\,\phi_n(s),
\]
with $\phi_n(s)=\sqrt{\frac{2}{L}}\cos(q_n s)$ and $q_n=\frac{n\pi}{L}.$
Projecting Eq.~\eqref{eq:u_eq_main} onto \(\phi_n\) gives independent mode equations,
\begin{equation}
\label{eq:mode_eq_main}
\zeta\,\dot a_n(t)+\kappa q_n^4 a_n(t)=\xi_n^{\rm th}(t)+\xi_n^{\rm act}(t),
\end{equation}
with relaxation rate $\omega_n={\kappa q_n^4}/{\zeta}$. The active projected force is
\[
\xi_n^{\rm act}(t)
=
-\int_0^L ds\,\phi_n'(s)\,F_{\rm act}(s,t),
\]
so that the derivative of the basis function contributes the characteristic factor \(q_n\).

Fourier transforming Eq.~\eqref{eq:mode_eq_main} in time gives
\[
\tilde a_n(\omega)
=
\frac{\tilde \xi_n^{\rm th}(\omega)+\tilde \xi_n^{\rm act}(\omega)}
{\kappa q_n^4-i\omega\zeta},
\qquad
|H_n(\omega)|^2
=
\frac{1}{\zeta^2(\omega^2+\omega_n^2)}.
\]
The mode variance therefore follows from the corresponding power spectra,
\[
\langle |a_n|^2\rangle
=
\int_{-\infty}^{\infty}\frac{d\omega}{2\pi}\,
|H_n(\omega)|^2
\big[S_n^{\rm th}(\omega)+S_n^{\rm act}(\omega)\big].
\]

The thermal contribution is standard. Since the tangent noise is the spatial derivative of the
thermal contour noise, one finds
\[
S_n^{\rm th}(\omega)=2k_B T\,\zeta\,q_n^2.
\]
For the active contribution, using the effective mode-space kernel
Eq.~\eqref{eq:mode_kernel_main}, the projected active-force correlation is
\[
\left\langle \xi_n^{\rm act}(t)\,\xi_n^{\rm act}(0)\right\rangle
=
q_n^2 D_0 e^{-q_n^2\sigma^2}e^{-|t|/\tau_{\rm eff}},
\]
whose Fourier transform is
\[
S_n^{\rm act}(\omega)
=
q_n^2 D_0 e^{-q_n^2\sigma^2}
\frac{2\tau_{\rm eff}}{1+\omega^2\tau_{\rm eff}^2}.
\]

Substituting the thermal and active spectra into the variance integral and evaluating the
frequency integral gives
\begin{equation}
\label{eq:mode_variance_main}
\langle |a_n|^2\rangle
=
\frac{1}{\kappa q_n^2}
\left[
k_B T
+
\frac{D_0\tau_{\rm eff}}{\zeta\,[1+\tau_{\rm eff}\omega_n]}
e^{-q_n^2\sigma^2}
\right].
\end{equation}
where $\omega_n=\kappa q_n^4/\zeta$ is the mode relaxation rate.
The derivation of the active-mode spectrum from the projected microscopic kernel, together with the
evaluation of the frequency integral leading to Eq.~\eqref{eq:mode_variance_main}, is given in
Appendices~\ref{app:mode_projection} and \ref{app:variance_derivation}.

Equation~\eqref{eq:mode_variance_main} can be interpreted as follows. The factor
$e^{-q_n^2\sigma^2}$ suppresses high modes because the active forcing is spatially coarse-grained over a finite range $\sigma$, while the denominator $1+\tau_{\rm eff}\omega_n$ expresses the competition between the active memory time and the intrinsic mode relaxation time. Slow modes with $\omega_n\tau_{\rm eff}\ll 1$ respond coherently to the active forcing, whereas fast modes with $\omega_n\tau_{\rm eff}\gg 1$ effectively average it out.

It is convenient to express the result in terms of a mode-dependent effective temperature,
defined by
\[
k_B T_n^{\rm eff}\equiv \kappa q_n^2 \langle |a_n|^2\rangle .
\]
This gives
\begin{equation}
\label{eq:Teff_main}
T_n^{\rm eff}
=
T+
\frac{D_0\tau_{\rm eff}}{k_B\zeta\,[1+\tau_{\rm eff}\kappa q_n^4/\zeta]}
e^{-q_n^2\sigma^2}.
\end{equation}
The polymer therefore does not experience a single effective temperature; instead, activity
generates a mode-resolved effective temperature profile that depends on both the temporal memory
and the spatial range of the active forcing.

\vskip 0.2cm
\noindent
\textit{Including a constant backbone tension.}
The weak-bending theory can be generalized in a straightforward way if the filament is subject to
a uniform tension $\cal{T}$. In that case, the deterministic mode stiffness is shifted according to
\[
\kappa q_n^4 \;\longrightarrow\; \kappa q_n^4+{\cal{T}} q_n^2,
\]
so that the relaxation rate becomes
\[
\omega_n({\cal{T}})=\frac{\kappa q_n^4+{\cal{T}} q_n^2}{\zeta}.
\]
The forcing statistics are unchanged, but the filtering by the polymer response is modified.
Accordingly, Eq.~\eqref{eq:mode_variance_main} becomes
\begin{equation}
\label{eq:mode_variance_with_tension}
\langle |a_n|^2\rangle
=
\frac{1}{\kappa q_n^2+\cal{T}}
\left[
k_B T
+
\frac{D_0\tau_{\rm eff}}{\zeta\,[1+\tau_{\rm eff}\omega_n({\cal{T}})]}
e^{-q_n^2\sigma^2}
\right].
\end{equation}
Tension therefore mainly suppresses the enhancement of the longest-wavelength modes, while leaving
the finite-range cutoff at high \(n\) unchanged.

\vskip 0.2cm
\noindent
\textit{Notes on comparison to simulations.}
Our simulations employ underdamped Langevin dynamics for both polymer beads and ABPs
(Sec.~\ref{sec:model}). The overdamped theory above is expected to hold for mode statistics on
timescales longer than the inertial relaxation time \(m/\gamma_t\), which is the regime probed by
the steady-state mode variances and effective temperatures reported in the Results.

\subsection{Radius of gyration from tangent modes}
\label{subsec:rg_main}

In the weak-bending, nearly inextensible regime we write
\[
\mathbf r(s)=x(s)\,\hat{\mathbf x}+\mathbf r_\perp(s),
\qquad
\mathbf u_\perp(s)\equiv \partial_s\mathbf r_\perp(s),
\]
and expand the transverse tangent field in free-end cosine modes, $\mathbf u_\perp(s)=\sum_{n\ge 1}\mathbf a_n\,\phi_n(s)$, with $\phi_n(s)=\sqrt{{2}/{L}}\cos(q_n s)$ and $q_n={n\pi}/{L}$.
The transverse coordinate follows from integration,
\[
\mathbf r_\perp(s)=\int_0^s ds'\,\mathbf u_\perp(s').
\]

\vskip 0.2cm
\noindent
\textit{Transverse contribution.}
The transverse part of the radius of gyration is
\[
R_{g,\perp}^2
=
\frac{1}{L}\int_0^L ds\,
\langle |\mathbf r_\perp(s)-\mathbf r_{\perp,\rm cm}|^2\rangle,
\]
where $\mathbf r_{\perp,\rm cm}=\frac{1}{L}\int_0^L ds\,\mathbf r_\perp(s).$
Assuming different tangent modes are uncorrelated,
\[
\langle \mathbf a_n\cdot \mathbf a_m\rangle
=
\delta_{nm}\,\langle |\mathbf a_n|^2\rangle,
\]
one obtains a weighted sum over mode variances,
\begin{equation}
\label{eq:rg_perp_weighted_main}
R_{g,\perp}^2
=
\sum_{n\ge 1}\frac{\langle |\mathbf a_n|^2\rangle}{Lq_n^{2}}\,W_n,
\qquad
W_n=
\begin{cases}
1, & n\ \text{even},\\[6pt]
1-\dfrac{8}{n^2\pi^2}, & n\ \text{odd}.
\end{cases}
\end{equation}

\vskip 0.2cm
\noindent
\textit{Longitudinal contribution.}
The longitudinal projection follows from local inextensibility,
\[
(\partial_s x)^2+|\mathbf u_\perp|^2=1,
\]
which gives, to leading order in the small-slope expansion,
\[
\partial_s x\simeq 1-\frac12|\mathbf u_\perp|^2.
\]
Thus
\[
x(s)=s-\epsilon(s),
\qquad
\epsilon(s)=\frac12\int_0^s ds'\,|\mathbf u_\perp(s')|^2.
\]
Keeping only the leading correction linear in \(\epsilon\) in
\[
R_{g,\parallel}^2=\frac{1}{L}\int_0^L ds\,\langle(x(s)-x_{\rm cm})^2\rangle,
\]
yields the mean-foreshortening estimate
\begin{equation}
\label{eq:rg_parallel_main}
R_{g,\parallel}^2
\simeq
\frac{L^2}{12}
-\frac{L}{12}\sum_{n\ge 1}\langle |\mathbf a_n|^2\rangle
+\frac{L}{4\pi^2}\sum_{n\ge 1}\frac{\langle |\mathbf a_n|^2\rangle}{n^2}.
\end{equation}

\vskip 0.2cm
\noindent
\textit{Total radius of gyration.}
To this order, the total radius of gyration is
\[
R_g^2 \simeq R_{g,\parallel}^2+R_{g,\perp}^2.
\]
This reconstruction assumes weak bending, a fixed contour length, and mode decorrelation at the two-point level. In particular, Eq.~\eqref{eq:rg_parallel_main} retains only the leading correction and neglects higher-order terms of order $O(|\mathbf u_\perp|^4)$. It is therefore expected to work best when the contour remains nearly inextensible. In regimes where the explicit active bath induces significant bond stretching, the theory should underestimate \(R_g\), since the activity-dependent increase of the contour length is not captured by the present fixed-contour weak-bending description. For comparison with simulations of a finite bead--spring chain, it is also useful to construct an exact finite-$N$ discrete reconstruction of $R_g$ in terms of bond-mode covariances which we discuss later. 


\section{Results}

\begin{figure*}[t] 
\centering 
\begin{tabular}{cc} 
\includegraphics[width=0.48\textwidth]{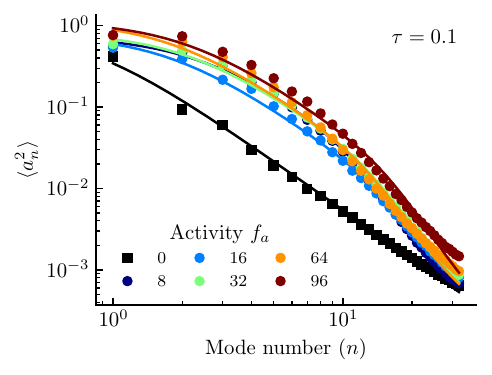} & \includegraphics[width=0.48\textwidth]{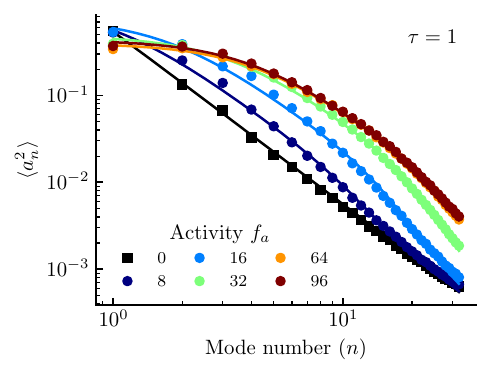} \\ \includegraphics[width=0.48\textwidth]{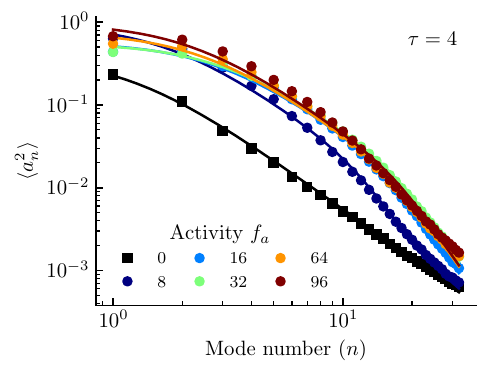} & \includegraphics[width=0.48\textwidth]{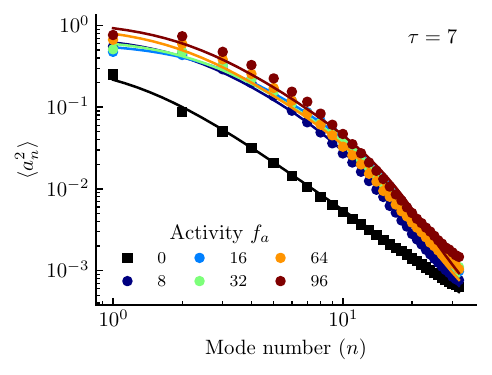} \\ \includegraphics[width=0.48\textwidth]{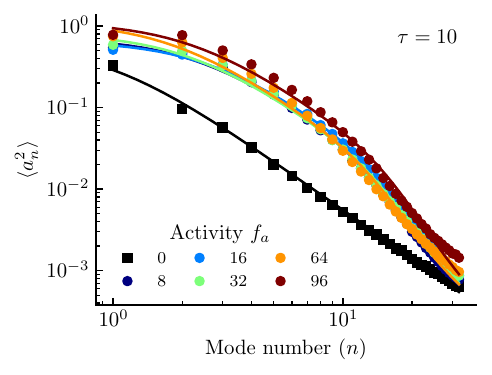} & \includegraphics[width=0.48\textwidth]{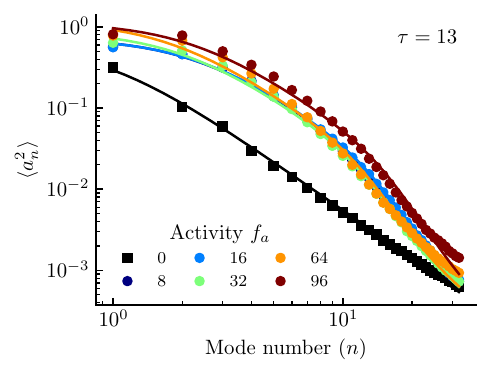} \end{tabular} 
\caption{ \textbf{Mode fluctuation spectrum as a function of activity at fixed persistence time.} Steady-state mode variances $\langle a_n^2\rangle$ are shown as a function of mode number $n$ for different activity strengths $f_a$ and for several fixed persistence times $\tau$. Symbols denote simulation data, and solid lines denote the theoretical fits. The passive reference spectrum ($f_a=0$) is shown for comparison in each panel. Increasing activity selectively enhances the low-mode sector, while the high-mode tail remains comparatively weakly affected, demonstrating that the active bath injects fluctuations non-uniformly across the spectrum. The theory captures both the overall magnitude and the crossover from activity-dominated low modes to nearly passive high modes over a broad range of parameters. } \label{fig:mode_fa} \end{figure*}

\subsection{Mode spectra: selective amplification by activity and persistence}

We first examine the steady-state tangent-mode fluctuations as functions of the activity strength $f_a$ and persistence time $\tau$. Since the effective theory is formulated directly at the level of mode amplitudes, the mode spectrum provides the most stringent test of the reduced description.

The spectra in Figs.~\ref{fig:mode_fa} and \ref{fig:mode_tau} show that the active bath modifies polymer fluctuations in a highly non-uniform manner. At fixed persistence time, increasing $f_a$ does not simply elevate all modes by the same amount. Instead, the enhancement is concentrated primarily in the lowest modes. The high-mode sector remains comparatively close to the passive case ($f_a = 0$). Thus, activity is selectively getting incorporated into the long-wavelength degrees of freedom; it is not a case of uniform heating.

A similar conclusion emerges when the persistence time is varied at fixed $f_a$. Increasing $\tau$ does not merely rescale the entire fluctuation spectrum upward. Rather, it shifts the spectral weight toward progressively lower mode numbers, so that the lowest modes are amplified most strongly while the high-mode tail remains only weakly affected. Persistence, therefore, acts as a spectral selection parameter: by increasing the temporal coherence of the forcing, it preferentially drives those modes whose intrinsic relaxation times are slow enough to remain correlated with the bath over a persistence window.

This behavior reflects the competition between the intrinsic relaxation time of mode $n$,
denoted $\tau_n$, and the active persistence time $\tau$. Modes with relaxation times
comparable to or longer than the active memory respond coherently and acquire a substantial nonequilibrium enhancement, whereas fast modes continue to average over the forcing and remain near-passive. The active bath, therefore, reorganizes the spectrum through a crossover between an activity-dominated low-mode sector and a weakly perturbed high-mode sector. As either $f_a$ or $\tau$ increases, this crossover shifts systematically toward lower mode number, indicating an increasing ability of the bath to drive large-scale polymer deformations.

\subsection{Comparison with theory and the limits of the reduced description}

For comparison with simulation data, we use a phenomenological fit
form motivated by Eq.~\eqref{eq:mode_variance_main}, with finite-chain and tension-like corrections
absorbed into effective fit parameters. The explicit form and fitting procedure are given in
Appendix~\ref{app:fitting}.

The theoretical fits superimposed on the mode spectra capture the main features of the simulation data across a broad range of parameters. In particular, the theory reproduces the selective enhancement of the low modes, the crossover from strongly active to weakly active sectors, and the systematic redistribution of spectral weight toward lower mode numbers as $\tau$ increases. This agreement supports the central premise of the theory: at the level of contour fluctuations, the explicit active bath can, to a good approximation, be captured by a temporally persistent, spatially coarse-grained forcing acting on mode fluctuations.

At the same time, the comparison also identifies the limits of this reduction. The most visible deviations occur in the strongest driving regimes, especially for the lowest modes at large $f_a$ and $\tau$, where the simulations exhibit somewhat stronger or broader enhancements than predicted by theory. These deviations indicate that under strong activity the polymer develops additional structural responses beyond those captured by the present weak-bending fixed-contour description. As we show below, the most important missing effect is activity-induced extensibility of the chain.

\begin{figure*}[t]
\centering
\begin{tabular}{cc}
\includegraphics[width=0.48\textwidth]{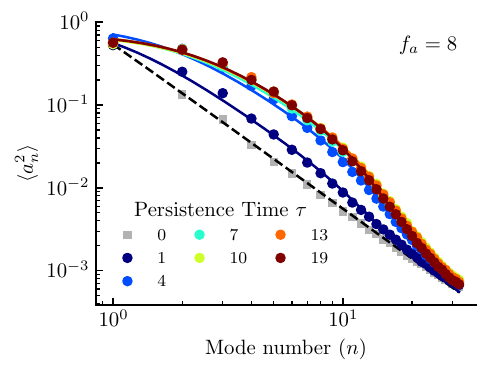} & \includegraphics[width=0.48\textwidth]{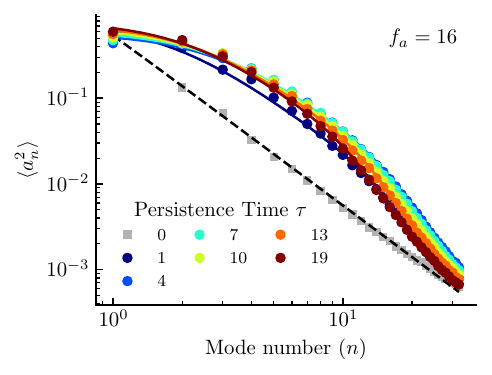} \\
\includegraphics[width=0.48\textwidth]{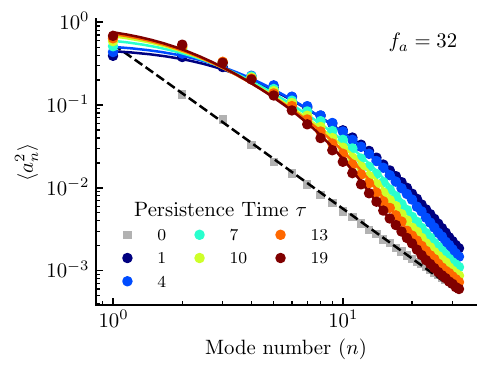} & \includegraphics[width=0.48\textwidth]{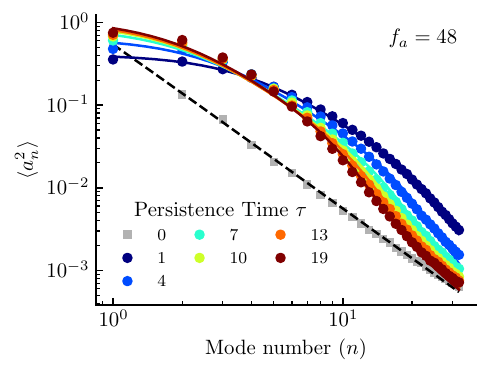} \\
\includegraphics[width=0.48\textwidth]{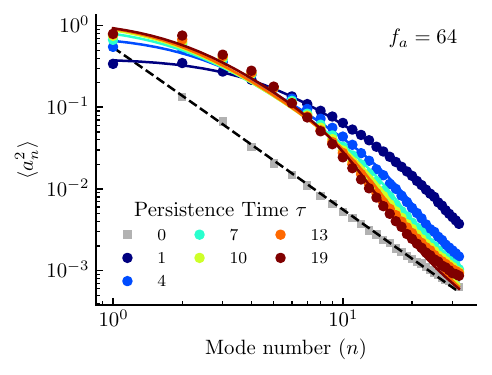} & \includegraphics[width=0.48\textwidth]{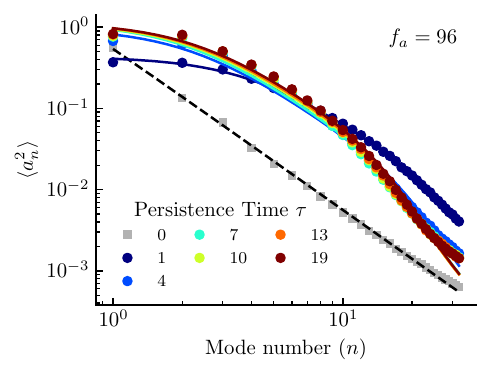}
\end{tabular}
\caption{
\textbf{Mode fluctuation spectrum as a function of persistence time at fixed activity.}
Steady-state mode variances $\langle a_n^2\rangle$ are shown as a function of mode number $n$ for different persistence times $\tau$ and for several fixed activity strengths $f_a$. Symbols denote simulation data and solid lines denote theoretical fits; the dashed line indicates the passive reference. Increasing persistence does not simply rescale the entire spectrum, but instead preferentially amplifies the lowest modes and shifts the crossover between active and weakly active sectors. This shows that the persistence time acts as a spectral selection parameter, making the active forcing increasingly coherent for slow polymer modes. The theory reproduces this spectral reorganization well, especially at intermediate and large mode numbers.
}
\label{fig:mode_tau}
\end{figure*}

A particularly important conclusion from Fig.~\ref{fig:mode_tau} is that the persistence time
cannot be interpreted as a simple noise-strength parameter. If the active bath merely raised an effective temperature, one would expect the entire mode spectrum to scale approximately uniformly. This is not what is observed. Instead, increasing $\tau$ preferentially amplifies the lowest modes and sharpens the separation between active and nearly passive sectors. A mode-dependent effective temperature can still be introduced as a convenient parametrization, but the underlying physics is more accurately described as a spectral crossover controlled by the interplay between active persistence and polymer relaxation.

\begin{figure*}[t!]
\centering
\begin{tabular}{cc}
\includegraphics[width=0.48\textwidth]{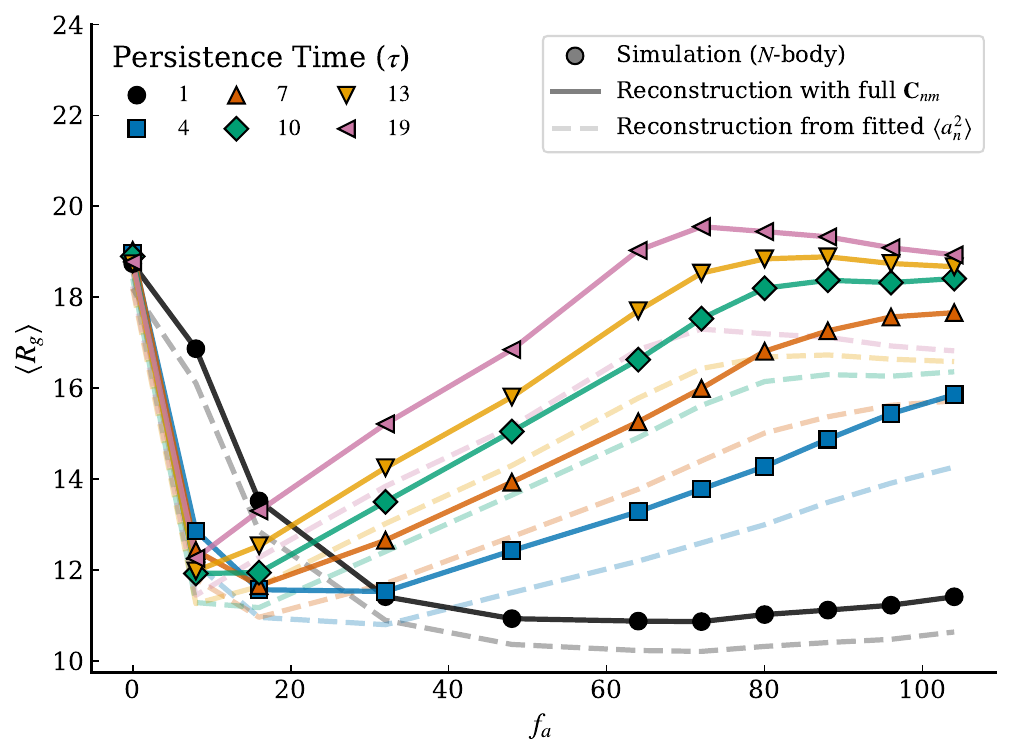} & \includegraphics[width=0.48\textwidth]{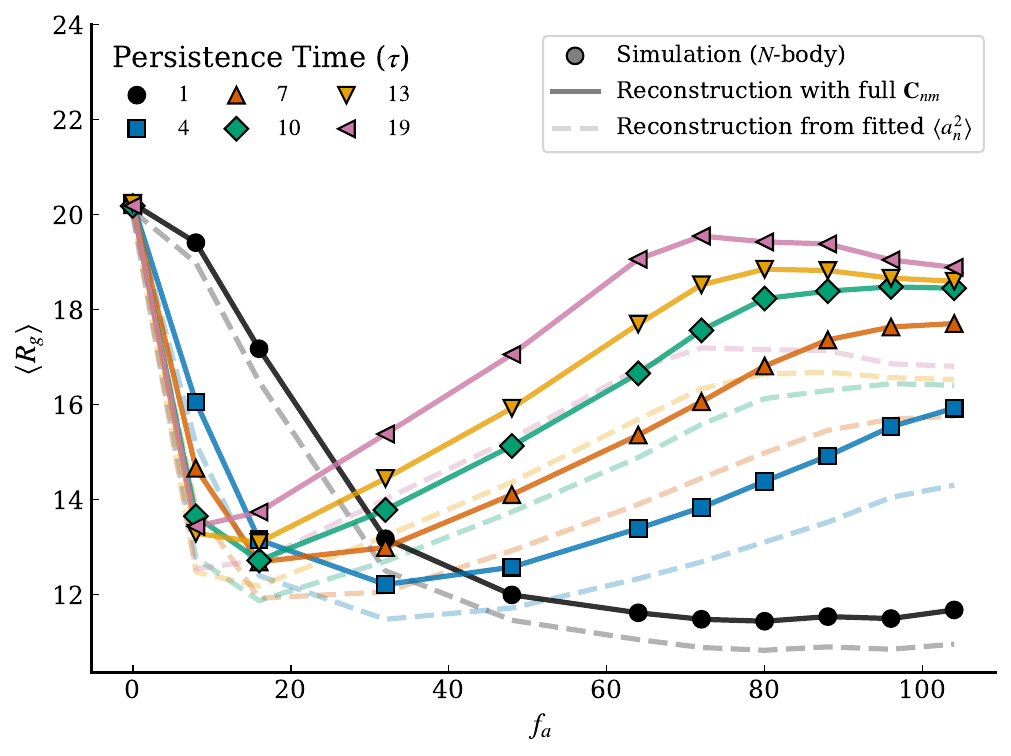}
\end{tabular}
\caption{
\textbf{Radius of gyration as a function of activity for two bending rigidities.}
Radius of gyration $R_g$ as a function of activity strength $f_a$ for several persistence times $\tau$, shown for (left) $\kappa=32k_{B}T$ and (right) $\kappa=128k_{B}T$. Symbols denote direct simulation measurements of $R_g$. The solid curves represent the exact full-covariance discrete reconstructions evaluated from the measured bond-mode covariance matrix (see Appendix \ref{app:finiteN_rg}), which capture the contributions of both mode variances and cross-mode couplings. The dashed curves are obtained from the tangent-series equation derived from microscopic details fitted with modes of bonds and the parameter space of the equation is obtained ($\mathcal{A},\mathcal{C},\mathcal{D},\mathcal{E}$) which assumes modes decoupling. The fit in the theory using the parameters for reconstruction captures the qualitative trend and the dependence on persistence time, but systematically underestimates the absolute magnitude of $R_g$, particularly at larger activity. This discrepancy indicates that, while the mode-level colored-noise description successfully captures the spectral redistribution of fluctuations, additional activity-induced geometric effects---most notably bond stretching and the associated renormalization of contour length---become important for global conformational measures.
}
\label{fig:rg}
\end{figure*}

\subsection{Radius of gyration: trends captured, magnitude underestimated}

We next compare the theory and simulations for the radius of gyration $R_g$, shown in
Fig.~\ref{fig:rg} for $\kappa=32k_{B}T$ and $\kappa=128k_{B}T$. In both cases the theory reproduces the qualitative trend that the polymer swells with increasing activity, and it also captures the dependence on persistence time $\tau$. However, the comparison reveals a clear and systematic shortfall: the theoretical prediction consistently underestimates the absolute magnitude of $R_g$, with the discrepancy becoming more pronounced at larger $f_a$ and larger $\tau$.

This mismatch is robust across both stiffnesses and therefore reflects a structural limitation of the present theory rather than a parameter-specific failure. It is important to distinguish this from the mode-spectrum comparison above. At the level of tangent-mode fluctuations, the reduced colored-noise description performs well and captures the principal spectral reorganization induced by activity. The disagreement in $R_g$, therefore, does not signal a breakdown of the mode-level picture itself. Instead, it indicates a combination of finite-chain geometric corrections, omitted mode cross-correlations, and activity-induced extensibility beyond the minimal continuum reconstruction.

\subsection{Activity-induced bond stretching and the limitation of the fixed-contour theory}

\begin{figure*}[t!]
\centering
\begin{tabular}{cc}
\includegraphics[width=0.48\textwidth]{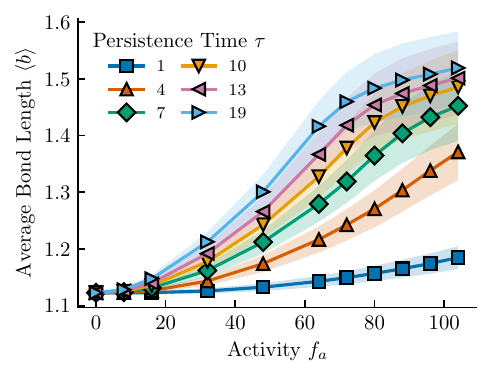} & \includegraphics[width=0.48\textwidth]{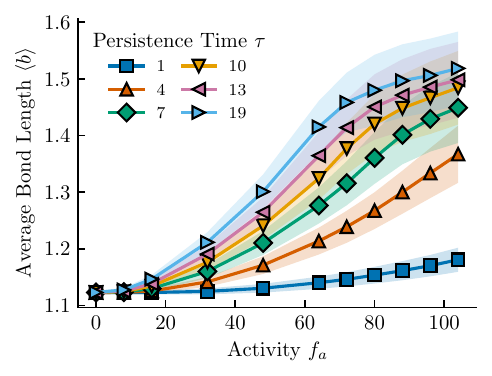}
\end{tabular}
\caption{
\textbf{Activity-induced increase of the mean bond length and steady-state bond fluctuations.}
Average bond length $\langle b\rangle$ (points) and bond-length fluctuation deviation (error bars) as a function of activity strength $f_a$ for different
persistence times $\tau$, shown for (left) $\kappa=32k_{B}T$ and (right) $\kappa=128k_{B}T$. The bond length increases significantly with activity in both case and the increase is stronger for larger persistence times. The activity-induced increase in bond length is found to be largely independent of the bending rigidity $\kappa$. This demonstrates that the polymer becomes effectively extensible in the explicit active bath, so that the contour length is renormalized by activity. The pronounced growth of $\langle b\rangle$ provides a natural explanation for the systematic underestimation of $R_g$ by the present weak-bending fixed-contour theory. The higher fluctuation also implied the tension helping in increasing cross mode coupling. 
}
\label{fig:bond}
\end{figure*}

The origin of the $R_g$ mismatch is clarified by the activity dependence of the mean bond length, shown in Fig.~\ref{fig:bond}. The average bond length increases significantly with $f_a$, and the increase becomes stronger as the persistence time is raised. This demonstrates that the chain becomes effectively extensible in the explicit active bath, so that the total contour length is itself renormalized by activity.

This observation is crucial because the present theoretical treatment of $R_g$ is built on a
weak-bending, fixed-contour reconstruction. Within that framework, activity modifies the
transverse mode amplitudes but leaves the backbone length unchanged. The simulations show that this assumption is violated under strong driving: activity not only amplifies bending fluctuations, but also stretches the chain longitudinally. Once this is recognized, the systematic underestimate of $R_g$ becomes natural. Since $R_g$ depends on both transverse shape fluctuations and overall contour extension, an increase in bond length necessarily pushes the simulation data above the fixed-contour prediction.

The discrepancy in $R_g$ should therefore be interpreted not as a failure of the spectral
colored-noise picture, but as evidence that explicit active baths generate additional geometric renormalization beyond mode-level forcing. A simple phenomenological improvement would be to replace the bare contour length by an activity-dependent effective contour length,
\begin{equation}
L_{\rm eff}(f_a,\tau)=(N-1)\,\langle b\rangle(f_a,\tau),
\end{equation}
and to use $L_{\rm eff}$ consistently in the mode wave numbers, contour reconstruction, and evaluation of global observables. Such a renormalized-length extension would preserve the successful mode-level description while incorporating the most direct structural effect seen in the explicit simulations.

\subsection{Exact finite-$N$ reconstruction of $R_g$ from bond-mode covariances}
To distinguish between errors arising from the continuum weak-bending reconstruction and those arising from genuinely missing physics, we next consider an exact finite-$N$ representation of the radius of gyration for the discrete bead--spring chain used in the simulations. Rather than working with a continuum tangent field, we decompose the bond vectors of the chain in a discrete cosine basis and reconstruct $R_g$ directly from their covariance matrix.

For a chain of $N+1$ monomers with bond vectors $\Delta \mathbf r_k \equiv \mathbf r_{k+1}-\mathbf r_k, \qquad k=0,\dots,N-1$, the radius of gyration may be written exactly as
\[
R_g^2
=
\frac{1}{(N+1)^2}\sum_{i=0}^{N}\sum_{j=i+1}^{N}
\left\langle |\mathbf r_j-\mathbf r_i|^2 \right\rangle .
\]
Expanding the bond vectors in a discrete cosine basis and summing them along the chain yields an
exact finite-$N$ reconstruction of the form
\begin{multline}
\label{eq:rg_discrete_cov}
R_g^2
=
W_{00}\,\langle \mathbf b_0\!\cdot\!\mathbf b_0\rangle
+\sum_{n\ge 1}W_{nn}\,\langle \mathbf b_n\!\cdot\!\mathbf b_n\rangle \\
+\sum_{n\ge 1}W_{0n}\,\langle \mathbf b_0\!\cdot\!\mathbf b_n\rangle
+\sum_{n<m}W_{nm}\,\langle \mathbf b_n\!\cdot\!\mathbf b_m\rangle,
\end{multline}
where the matrix of geometric weights $W_{nm}$ depends only on the finite chain length and on the
chosen discrete basis. The full derivation and explicit finite-$N$ weights are given in
Appendix~\ref{app:finiteN_rg}.

The importance of Eq.~\eqref{eq:rg_discrete_cov} is that it separates two distinct approximations that are combined in the continuum weak-bending reconstruction of Sec.~III.E. First, the continuum treatment replaces the exact finite-chain geometry by its large-$N$ limit. Second, the leading-order reconstruction retains only a restricted subset of mode covariances. By contrast, Eq.~\eqref{eq:rg_discrete_cov} is an exact kinematic identity for the finite chain once the full covariance matrix of discrete bond modes is included.

When evaluated using the full simulation-measured covariance matrix, the discrete reconstruction reproduces the directly measured $R_g$ essentially exactly. This shows that the discrepancy seen in Fig.~\ref{fig:rg} is not caused by an inconsistency in the mode description itself, but by the combination of (i) continuum weak-bending geometric approximations and (ii) omission of cross-correlations and extension-related contributions in the leading-order fixed-contour reconstruction. 


\section{Discussion and Conclusion}

The central goal of this work was to understand how an explicit active environment is encoded in the internal fluctuation spectrum of a semiflexible polymer. We have combined an effective theory for mode-resolved forcing with simulations of both explicit-bath and implicit-noise models. We arrive at a simple physical conclusion: the active bath does not act as a featureless source of enhanced agitation, but reorganizes polymer fluctuations spectrally. The extent of enhancement depends strongly on the competition between the bath persistence time and the hierarchy of polymer relaxation times.

We find strong mode selectivity of the active response.
At fixed persistence time, increasing the active force $f_a$ preferentially amplifies the
longest-wavelength modes, while the high-mode sector remains comparatively close to passive behavior. Similarly, at fixed activity, increasing the persistence time $\tau$ does not simply rescale the entire spectrum, but shifts the spectral crossover so that progressively lower modes become dominant. The persistence time therefore acts as a mode-selection parameter: slow modes respond coherently to persistent forcing, whereas fast modes continue to average over it. This is the basic reason why the notion of a single effective temperature is inadequate at the level of internal polymer dynamics.

At the same time, the results show that a reduced colored-noise description remains highly useful. The mode spectra are reproduced well in a broad range of $f_a$ and $\tau$. This includes selective enhancement of low modes, the crossover from activity-dominated to weakly perturbed sectors, and the redistribution of spectral weight with persistence. In this sense, much of the nonequilibrium complexity of the explicit bath can be compressed into a small number of effective parameters controlling the temporal memory and spatial coarse-graining of the forcing. This is one of the main positive messages of the paper.


However, the comparison with simulations also makes it clear that this reduction is not complete. The largest deviations in the mode spectra occur at the strongest drives and longest persistence times, especially in the lowest modes. More importantly, the theory systematically underestimates the radius of gyration $R_g$ for both $\kappa=32k_{B}T$ and $\kappa=128k_{B}T$, although it captures the qualitative trend of increasing the size of the polymer with activity. 


The bond-length data provide the most direct explanation of this mismatch. In explicit bath simulations, the mean bond length increases significantly with activity and more strongly at longer persistence times. The polymer, therefore, becomes effectively extensible under strong driving. This effect is absent from the present reconstruction of $R_g$, which assumes a fixed contour length and attributes the nonequilibrium response entirely to transverse mode fluctuations. 
An exact finite-$N$ reconstruction of $R_g$ from bond-mode covariances further confirms that the mode representation itself remains complete at the discrete level. The residual mismatch of the minimal continuum theory, therefore, originates from the combined effect of continuum geometric approximations, neglected cross-correlations,  and activity-induced extensibility.

This observation suggests a clear route for improving the theory. A minimal phenomenological extension would be to replace the bare contour length by an activity-dependent effective contour length. Such a modification would preserve the successful mode-resolved forcing picture while incorporating the most direct geometric renormalization seen in the explicit simulations. 


A second broader implication of our results concerns the relationship between explicit and implicit active baths. The comparison carried out here shows that an implicit colored-noise model can reproduce the leading spectral trends of the explicit bath very well, especially for the large-scale modes that dominate global conformational fluctuations. This qualitative similarity is visible in the normalized $R_g$ landscapes of the two models shown in Appendix~\ref{app:implicit_compare}. At the same time, the comparison also clarifies the limits of such a mapping. Finite particle size, excluded-volume collisions, crowding, and activity-induced bond stretching leave signatures that are not automatically encoded in a minimal implicit-noise description. 

The present results also sharpen the meaning of effective temperature in active matter. A
mode-dependent effective temperature can be introduced as a compact representation of the measured mode amplitudes. However, as we have seen, the polymer filters the active bath through its own relaxation spectrum. Different wavelengths therefore experience different effective forcing strengths and different degrees of temporal coherence. From this point of view, the polymer acts as a multiscale probe of active matter, resolving the temporal and spatial structure of nonequilibrium forcing through its mode spectrum.

Recent experiments on chromatin dynamics point to a related physical picture. In particular,
single-gene and genome-scale motion measurements in live human cells have shown that the coupling
between localized transcription-dependent activity and surrounding chromatin motion depends
sensitively on the local chromatin compaction~\cite{chu2024transcription,zidovska2020self}. More compact regions appear to suppress or localize the transmission of activity, whereas less compact regions exhibit stronger coherent coupling to larger-scale motions. Although the biological setting is far more complex than the minimal polymer-bath model studied here, these observations are broadly consistent with the idea that nonequilibrium forcing is filtered by the internal mechanical and relaxation structure of
the medium. 

Several extensions naturally follow from this work. The most immediate is to incorporate
activity-induced extensibility into the reduced theory, thereby improving the description of
global size observables such as $R_g$. A second direction is to analyze spectral energy injection and transfer more explicitly to determine how activity is distributed and dissipated across modes in the nonequilibrium steady state. It would also be interesting to generalize the framework to situations with tension, confinement, or anisotropic active forcing, where the balance between bending relaxation and active persistence may be qualitatively altered. More broadly, the present approach suggests that embedded polymers could be used as spectroscopic probes of complex active media: by measuring their mode spectra, one may infer the characteristic temporal and spatial scales of the underlying activity.

In conclusion, the response of a semiflexible polymer in an active bath is fundamentally spectral. The active environment does not simply ``heat'' the polymer uniformly; instead, it couples selectively to modes according to the interplay between bath persistence and polymer relaxation. A reduced colored-noise theory captures this mode-level physics with notable success and explains the main spectral trends seen in explicit-bath simulations. At the same time, the systematic underestimate of $R_g$ and the pronounced increase in bond length show that explicit active baths also generate geometric renormalizations, especially activity-induced extensibility, that lie beyond the present weak-bending fixed-contour description. 

\section{Conflicts of Interests}
There are no conflicts to declare.

\bibliography{references}

\appendix

\section{{Mapping of simulation parameter }$\alpha$ {to persistence time $\tau$}}
\label{app:angmom}

In LAMMPS, the rotational Langevin thermostat for ABPs is controlled by a parameter (internally  \texttt{angmom}, denoted $\alpha$ here), defined as the ratio of translational to rotational inertial relaxation times~\cite{dias2021molecular}:
\begin{equation}
\alpha = \frac{\tau_t}{\tau_r} = \frac{m/\gamma_t}{I/\gamma_r},
\end{equation}
where $\gamma_t$ and $\gamma_r$ are the translational and rotational friction coefficients, and $I$ is the moment of inertia. Using the fluctuation-dissipation relation $\gamma_r = k_BT/D_r$, where $D_r$ is the rotational diffusion rate, the physical orientational persistence time used throughout this work is defined as $\tau \equiv 1/D_r$. Rearranging the expression for $\alpha$, we find that $\tau$ is directly proportional to the simulation parameter:
\begin{equation}
\label{eq:taur_angmom_new}
\tau = \frac{\gamma_t\,I}{m\,k_BT} \times \alpha.
\end{equation}
For a sphere of diameter $\sigma$, the moment of inertia is $I = m\sigma^2/10$. With our simulation parameters ($\gamma_t=10$, $\sigma=1$, $k_BT=1$, $m=2/3$), the prefactor evaluates to unity, yielding $\tau = \alpha$ exactly. Throughout this work, we use the symbol $\tau$ to represent this physical persistence parameter.

\section{Numerical mode analysis and parameter extraction}
\label{app:fitting}

To rigorously bridge the continuum theory with the discrete $N$-body simulations, we perform a systematic modal decomposition of the polymer configurations. For each parameter set $(f_a, \tau)$, observables are time-averaged over the steady-state portion of the trajectory, taking the final 30\% of the $10^8$ timesteps to ensure transient initializations have decayed.

\textit{Mode extraction and covariance matrix: }
From the instantaneous simulation coordinates of the \(N\) beads, we construct the \(N-1\) discrete
bond vectors $\Delta \mathbf r_i = \mathbf r_{i+1}-\mathbf r_i$.
These bond vectors are projected into a discrete cosine basis to yield the instantaneous bond-mode
amplitudes $\mathbf b_n$ (see Appendix~\ref{app:finiteN_rg}). We then compute the full time-averaged covariance matrix $C_{nm}=\langle \mathbf b_n\cdot \mathbf b_m\rangle$. To map these spectral fluctuations back to a macroscopic structural observable without invoking a continuum weak-bending approximation, the mean-squared radius of gyration is evaluated in the exact finite-\(N\) discrete basis as
\begin{equation}
\langle R_g^2\rangle=\sum_{n,m} C_{nm}W_{nm},
\end{equation}
where \(W_{nm}\) is the discrete geometric weight matrix determined entirely by the finite-chain
projection. When the measured covariance matrix is inserted, this full-matrix evaluation reproduces
the exact simulated \(\langle R_g^2\rangle\) and therefore serves as a benchmark for approximate
theoretical reconstructions.

\textit{Spectral fitting procedure: }
Theoretical predictions for the steady-state mode variances, identified with the diagonal elements
\(C_{nn}\equiv \langle |\mathbf a_n|^2\rangle\), are fitted to the simulation data using a
two-stage non-linear least-squares procedure implemented with \texttt{scipy.optimize.curve\_fit}.
To prevent the fit from being dominated entirely by the most strongly excited low-\(n\) modes, we
apply relative weights proportional to \(1/\langle |\mathbf a_n|^2\rangle\) over the fitted range
\(n=1,\dots,32\).

In the first stage, the passive spectrum (\(f_a=0\)) is fitted to the equilibrium form
\begin{equation}
\langle |a_n|^2\rangle = \frac{\mathcal A}{n^2+\mathcal C_p},
\end{equation}
yielding the passive fluctuation amplitude \(\mathcal A\) and an effective offset \(\mathcal C_p\)
that absorbs finite-chain and discretization corrections.

In the second stage, the passive amplitude \(\mathcal A\) is held fixed, and the active spectra are
fitted using a phenomenological form motivated by the tension-generalized mode-variance theory:
\begin{equation}
\label{eq:fit_form_app}
\langle |a_n|^2\rangle
=
\frac{\mathcal A}{n^2+\mathcal C}
\left[
1+\frac{\mathcal D}{1+\mathcal E^2(n^4+\mathcal C n^2)}
\right].
\end{equation}
Here, $\mathcal C$ plays the role of an effective tension-like parameter, while $\mathcal D$
controls the strength of the active enhancement. 
Fits are performed independently for each $(f_a,\tau)$ pair. To improve convergence---especially at high activity, where the low-mode
enhancement becomes pronounced---we sweep through increasing $f_a$, using the converged parameters from the previous activity step as the initial guess for the next.

\section{Qualitative comparison of explicit and implicit active baths}
\label{app:implicit_compare}

\begin{figure*}[t]
\centering
\includegraphics[width=\textwidth]{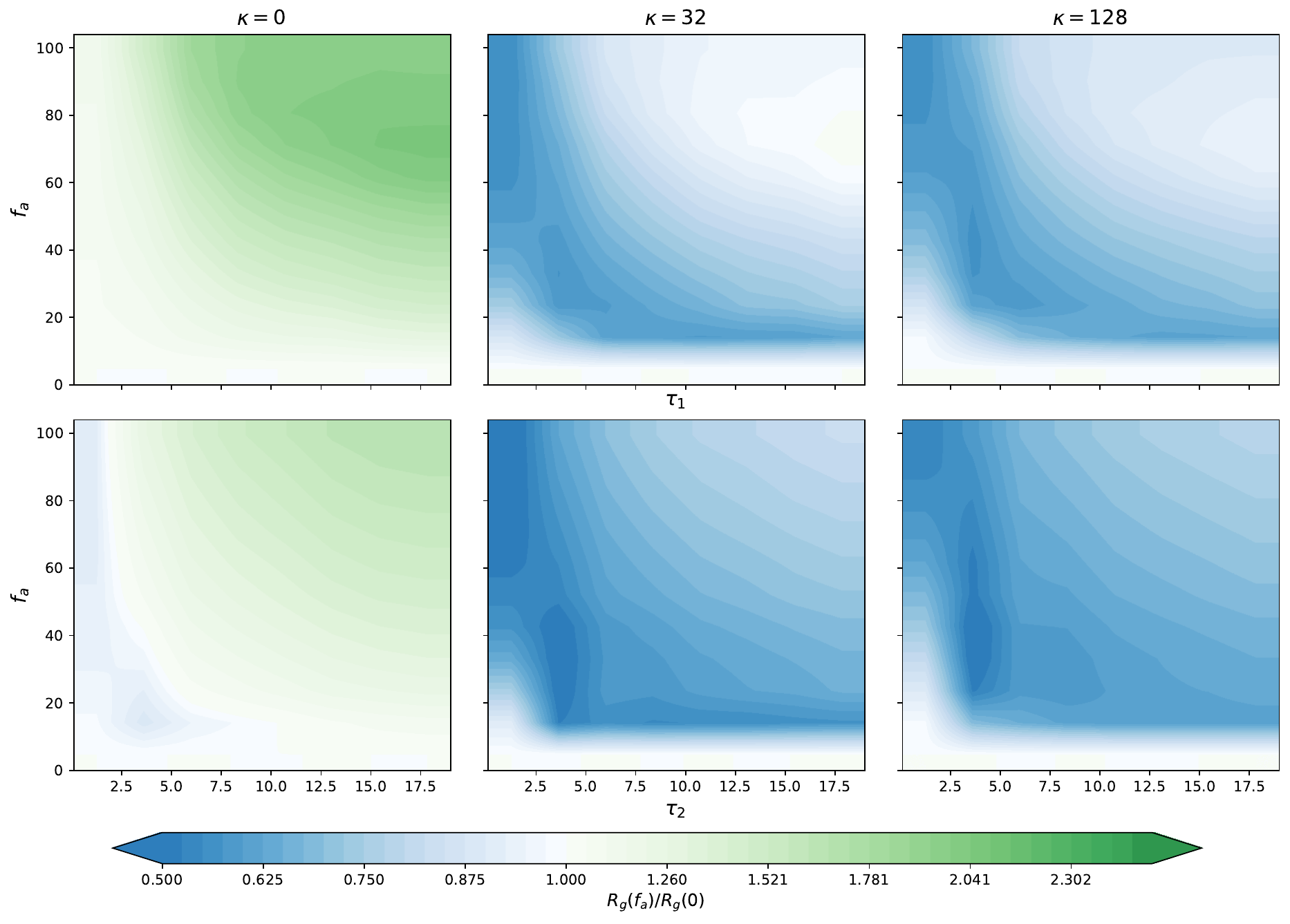}
\caption{
\textbf{Qualitative comparison of polymer swelling in explicit and implicit active baths.}
Heatmaps of the normalized radius of gyration, $R_g(f_a)/R_g(0)$, in the $(f_a,\tau)$ plane for
the explicit active bath (top row) and the implicit Ornstein--Uhlenbeck forcing model (bottom row),
shown for three representative values of the bending rigidity. The two models display closely
similar conformational trends over parameter space, indicating that the implicit OU description
captures the leading large-scale swelling phenomenology of the explicit bath. The analytical fits
in the main text, however, are carried out only for the explicit-bath model, and no strict
one-to-one microscopic parameter mapping is imposed between the two descriptions.
}
\label{fig:heatmap_compare}
\end{figure*}

To assess how closely a minimal colored-noise description reproduces the broad conformational
response of the explicit active bath, we compare the normalized radius of gyration
$R_g(f_a)/R_g(0)$ over the corresponding activity--persistence parameter planes for the two
models. Figure~\ref{fig:heatmap_compare} shows the resulting heatmaps for the explicit active bath
(top row) and the implicit Ornstein--Uhlenbeck forcing model (bottom row), for three representative
values of the bending rigidity.

The main observation is that the two models display closely similar large-scale conformational
trends. In both cases, polymer swelling is weak at small activity and short persistence, and
becomes progressively stronger as either the forcing amplitude or the persistence time is increased.
This qualitative similarity indicates that a temporally persistent colored-noise forcing captures
the leading large-scale swelling phenomenology of the explicit active bath over a broad region of
parameter space.

At the same time, this comparison is intended only as a qualitative benchmark. In the main text,
the analytical fitting is carried out exclusively for the explicit-bath model, and no strict
one-to-one microscopic mapping between the parameters of the explicit and implicit descriptions is
imposed. The heatmaps should therefore be interpreted as evidence that the implicit OU model
reproduces the broad conformational landscape of the explicit bath, rather than as proof of
quantitative equivalence between the two models.

\section{Microscopic force correlator for an explicit ABP bath}
\label{app:forcecorr}

In this appendix we derive the real-space force correlator used in the main text. The starting point
is the active force density exerted by the bath on the polymer contour,
\[
\mathbf F_{\rm act}(s,t)=\sum_{i=1}^{N_a}\mathbf f\!\left(\mathbf R_i(t)-\mathbf r(s,t)\right),
\]
with $\mathbf f(\mathbf x)\equiv -\nabla V(|\mathbf x|)$. For a homogeneous dilute bath, the mean force vanishes by isotropy, and the leading statistical
object is the connected correlator
\[
C_{\alpha\beta}(s,t;s',t')
=
\left\langle F_{{\rm act},\alpha}(s,t)\,F_{{\rm act},\beta}(s',t')\right\rangle_c.
\]
Neglecting bath--bath correlations beyond the uniform density $\rho_0$ and retaining only the single-particle contribution gives
\begin{multline}
\label{eq:prop}
C_{\alpha\beta}(s,t;s',t')
\simeq
\rho_0
\int d^d\mathbf R \int d^d\mathbf R'\,
f_\alpha(\mathbf R-\mathbf r(s,t)) \\
f_\beta(\mathbf R'-\mathbf r(s',t'))
P(\mathbf R',t'|\mathbf R,t),
\end{multline}
where $P(\mathbf R',t'|\mathbf R,t)$ is the single-particle propagator of a bath particle. The origin of this expression is straightforward. Since
$F_{{\rm act},\alpha}(s,t)=\sum_{i=1}^{N_a} f_\alpha\!\left(\mathbf R_i(t)-\mathbf r(s,t)\right),$ the two-point correlator involves a double sum over bath-particle labels:
\begin{multline}
\nonumber
\left\langle F_{{\rm act},\alpha}(s,t)\,F_{{\rm act},\beta}(s',t')\right\rangle
=\\
\sum_{i,j}
\left\langle
f_\alpha\!\left(\mathbf R_i(t)-\mathbf r(s,t)\right)
f_\beta\!\left(\mathbf R_j(t')-\mathbf r(s',t')\right)
\right\rangle .
\end{multline}
In a homogeneous isotropic bath, the mean force vanishes, so the connected and ordinary second
moments coincide. Under the dilute-bath approximation, correlations between distinct bath particles ($i\neq j$) contribute only through the disconnected part and are therefore neglected. The
leading contribution then comes from the same particle appearing at both times $(i=j)$.
Writing the corresponding two-time single-particle probability density as $p_2(\mathbf R,t;\mathbf R',t')=p_1(\mathbf R,t)P(\mathbf R',t'|\mathbf R,t)$, and using the uniform one-particle density $N_a p_1=\rho_0$, gives Eq.~\ref{eq:prop}. Thus the force correlator is the same-particle contribution: a bath particle exerts a force on the polymer at $(s,t)$, propagates according to $P(\mathbf R',t'|\mathbf R,t)$, and exerts a second force at $(s',t')$.


Introducing the Fourier transform
\[
f_\alpha(\mathbf x)=\int \frac{d^d\mathbf k}{(2\pi)^d}\,\tilde f_\alpha(\mathbf k)\,
e^{i\mathbf k\cdot \mathbf x},
\]
and approximating the propagator by its free-space form
$P(\mathbf R',t'|\mathbf R,t)\approx P(\mathbf R'-\mathbf R,\Delta t)$ with
$\Delta t=t'-t$, the correlator becomes
\begin{multline}
\nonumber
C_{\alpha\beta}(s,t;s',t')
=
\rho_0
\int \frac{d^d\mathbf k}{(2\pi)^d}
\int \frac{d^d\mathbf k'}{(2\pi)^d}\,
\tilde f_\alpha(\mathbf k)\tilde f_\beta(\mathbf k')\,\\
e^{-i\mathbf k\cdot \mathbf r(s,t)}
e^{-i\mathbf k'\cdot \mathbf r(s',t')}
I(\mathbf k,\mathbf k';\Delta t),
\end{multline}
with
\[
I(\mathbf k,\mathbf k';\Delta t)
=
\int d^d\mathbf R \int d^d\mathbf R'\,
e^{i\mathbf k\cdot \mathbf R}e^{i\mathbf k'\cdot \mathbf R'}\,
P(\mathbf R'-\mathbf R,\Delta t).
\]
Introduce the relative coordinate $\Delta \mathbf R=\mathbf R'-\mathbf R$ and the center-of-mass
coordinate $\mathbf X=\mathbf R$. Then
\begin{multline}
\nonumber
I(\mathbf k,\mathbf k';\Delta t)
=
\int d^d\mathbf X\,e^{i(\mathbf k+\mathbf k')\cdot \mathbf X}\\
\int d^d(\Delta \mathbf R)\,e^{i\mathbf k'\cdot \Delta \mathbf R}
P(\Delta \mathbf R,\Delta t).
\end{multline}
The integral over $\mathbf X$ yields
\[
\int d^d\mathbf X\,e^{i(\mathbf k+\mathbf k')\cdot \mathbf X}
=
(2\pi)^d\delta(\mathbf k+\mathbf k').
\]
Hence
\[
I(\mathbf k,\mathbf k';\Delta t)
=
(2\pi)^d\delta(\mathbf k+\mathbf k')\,
\left\langle e^{i\mathbf k'\cdot \Delta \mathbf R(\Delta t)}\right\rangle,
\]
where the average is over the displacement distribution of a single ABP. Substituting this back
and integrating over $\mathbf k'$ gives
\begin{multline}
\label{eq:Ck_appendix}
C_{\alpha\beta}(s,t;s',t')
=
\rho_0\int\frac{d^d\mathbf k}{(2\pi)^d}\,
\tilde f_\alpha(\mathbf k)\tilde f_\beta(-\mathbf k)\,
e^{i\mathbf k\cdot \Delta \mathbf r}\,\\
\left\langle e^{-i\mathbf k\cdot \Delta \mathbf R(\Delta t)}\right\rangle,
\end{multline}
with $\Delta \mathbf r=\mathbf r(s',t')-\mathbf r(s,t)$.

\vskip 0.2cm
\noindent
The characteristic function in Eq.~\eqref{eq:Ck_appendix} is approximated by truncating the
cumulant expansion at second order:
\[
\left\langle e^{-i\mathbf k\cdot \Delta \mathbf R(\Delta t)}\right\rangle
\simeq
\exp\!\left[-\frac12 k_\mu k_\nu
\left\langle \Delta R_\mu(\Delta t)\Delta R_\nu(\Delta t)\right\rangle\right].
\]
For an isotropic bath,
\[
\left\langle \Delta R_\mu(\Delta t)\Delta R_\nu(\Delta t)\right\rangle
=
2S(\Delta t)\,\delta_{\mu\nu},
\]
so that
\[
\left\langle e^{-i\mathbf k\cdot \Delta \mathbf R(\Delta t)}\right\rangle
\simeq
e^{-k^2S(\Delta t)}.
\]
Substitution into Eq.~\eqref{eq:Ck_appendix} yields
\begin{equation}
\label{eq:Ck_gaussian_appendix}
C_{\alpha\beta}(\Delta \mathbf r,\Delta t)
=
\rho_0\int\frac{d^d\mathbf k}{(2\pi)^d}\,
\tilde f_\alpha(\mathbf k)\tilde f_\beta(-\mathbf k)\,
e^{i\mathbf k\cdot \Delta \mathbf r}\,
e^{-k^2S(\Delta t)}.
\end{equation}

\vskip 0.2cm
\noindent
For an active Brownian particle,
\[
\Delta \mathbf R(\Delta t)
=
v_0\int_0^{\Delta t}\hat{\mathbf u}(s)\,ds
+
\sqrt{2D_t}\,\big[\mathbf W(t+\Delta t)-\mathbf W(t)\big].
\]
Using the isotropic orientation correlator
\[
\langle \hat u_i(s)\hat u_j(s')\rangle
=
\frac{\delta_{ij}}{d}\,e^{-|s-s'|/\tau},
\]
one finds
\begin{multline}
\nonumber
\langle \Delta R_i(\Delta t)\Delta R_j(\Delta t)\rangle
=\\
\delta_{ij}\left[
2D_t\,\Delta t
+\frac{2v_0^2}{d}
\left(
\tau\Delta t-\tau^2[1-e^{-|\Delta t|/\tau}]
\right)
\right].
\end{multline}
It is therefore convenient to define
\begin{equation}
\label{eq:S_appendix}
S(\Delta t)
=
D_t\,|\Delta t|
+
\frac{v_0^2\tau}{d}
\left(
|\Delta t|-\tau[1-e^{-|\Delta t|/\tau}]
\right).
\end{equation}
At short times, \(S(\Delta t)\sim D_t|\Delta t|+\frac{v_0^2}{2d}\Delta t^2\), while at long times
it crosses over to the diffusive form
\[
S(\Delta t)\simeq D_{\rm eff}|\Delta t|,
\qquad
D_{\rm eff}=D_t+\frac{v_0^2\tau}{d}.
\]

\vskip 0.2cm
\noindent
For the Gaussian potential $ V(r)=V_0 e^{-r^2/(2\sigma^2)},$ the force Fourier transform satisfies
\[
\tilde f_\alpha(\mathbf k)\tilde f_\beta(-\mathbf k)
=
k_\alpha k_\beta\,V_0^2\,(2\pi\sigma^2)^d\,e^{-\sigma^2k^2}.
\]
Equation~\eqref{eq:Ck_gaussian_appendix} therefore becomes
\[
C_{\alpha\beta}(\Delta \mathbf r,\Delta t)
=
\rho_0V_0^2\sigma^{2d}
\int d^d\mathbf k\,
k_\alpha k_\beta\,
e^{-A(\Delta t)k^2+i\mathbf k\cdot \Delta \mathbf r},
\]
where
\[
A(\Delta t)=\sigma^2+S(\Delta t).
\]
Using the identity
\[
\int d^d\mathbf k\,
k_\alpha k_\beta e^{-Ak^2+i\mathbf k\cdot \Delta \mathbf r}
=
-\partial_{\Delta r_\alpha}\partial_{\Delta r_\beta}
\int d^d\mathbf k\,e^{-Ak^2+i\mathbf k\cdot \Delta \mathbf r},
\]
together with
\[
\int d^d\mathbf k\,e^{-Ak^2+i\mathbf k\cdot \Delta \mathbf r}
=
\left(\frac{\pi}{A}\right)^{d/2}
\exp\!\left[-\frac{|\Delta \mathbf r|^2}{4A}\right],
\]
one finds after differentiation
\begin{multline}
\label{eq:Creal_appendix}
C_{\alpha\beta}(\Delta \mathbf r,\Delta t)
=
\rho_0V_0^2\sigma^{2d}
\left(\frac{\pi}{A(\Delta t)}\right)^{d/2}\\
\left[
\frac{\delta_{\alpha\beta}}{2A(\Delta t)}
-\frac{\Delta r_\alpha\Delta r_\beta}{4A(\Delta t)^2}
\right]
\exp\!\left[-\frac{|\Delta \mathbf r|^2}{4A(\Delta t)}\right].
\end{multline}
This is the real-space force correlator quoted in the main text.

\section{Projection of the microscopic force kernel onto tangent modes}
\label{app:mode_projection}

In this appendix we project the microscopic force correlator onto tangent modes. This provides the link between the explicit-bath correlator in Eq.~\eqref{eq:Creal_appendix} and the effective mode-space forcing used in the main text. In the weak-bending regime, the polymer is nearly straight on the scales relevant to the force
kernel. We therefore approximate
\[
|\mathbf r(s)-\mathbf r(s')|^2 \simeq (s-s')^2 .
\]
Then the isotropic part of Eq.~\eqref{eq:Creal_appendix} depends only on
\(\Delta s=s-s'\), and the force correlator becomes approximately translationally invariant along
the contour:
\[
C(s,t;s',t')\simeq C(\Delta s,\Delta t),
\]
with
\[
C(\Delta s,\Delta t)
=
\mathcal C_0\,A(\Delta t)^{-p}
\exp\!\left[-\frac{\Delta s^2}{4A(\Delta t)}\right].
\]
Here \(\mathcal C_0\) absorbs the prefactors in Eq.~\eqref{eq:Creal_appendix}, and \(p\) denotes
the effective power of \(A(\Delta t)\) multiplying the Gaussian factor. The precise value of \(p\)
is not essential for the projection; what matters is that the spatial kernel remains Gaussian with
a width controlled by \(A(\Delta t)\).


The active tangent force on mode \(n\) is
\[
\xi_n^{u,{\rm act}}(t)
=
-\int_0^L ds\,\phi_n'(s)\,F_{\rm act}(s,t),
\]
where
\[
\phi_n(s)=\sqrt{\frac{2}{L}}\cos(q_n s),
\qquad
\phi_n'(s)=-q_n\sqrt{\frac{2}{L}}\sin(q_n s).
\]
Its correlator is therefore
\begin{multline}
\nonumber
\left\langle \xi_n^{u,{\rm act}}(t)\xi_m^{u,{\rm act}}(t')\right\rangle
=\\
\int_0^L ds\int_0^L ds'\,
\phi_n'(s)\phi_m'(s')\,C(s-s',\Delta t).
\end{multline}
Substituting the explicit form of \(\phi_n'\) gives
\begin{multline}
\nonumber
\left\langle \xi_n^{u,{\rm act}}(t)\xi_m^{u,{\rm act}}(t')\right\rangle
=\\
\frac{2q_nq_m}{L}
\int_0^L ds\int_0^L ds'\,
\sin(q_ns)\sin(q_ms')\,C(s-s',\Delta t).
\end{multline}


Since \(C(s-s',\Delta t)\) is a convolution kernel, it is diagonalized approximately by Fourier
modes. It is therefore convenient to introduce the spatial Fourier transform
\[
\widetilde C(q,\Delta t)
=
\int_{-\infty}^{\infty} d(\Delta s)\,
e^{iq\Delta s}\,C(\Delta s,\Delta t).
\]
Using the Gaussian form of the kernel,
\[
C(\Delta s,\Delta t)
=
\mathcal C_0\,A(\Delta t)^{-p}
\exp\!\left[-\frac{\Delta s^2}{4A(\Delta t)}\right],
\]
we obtain
\[
\widetilde C(q,\Delta t)
=
\mathcal C_0\,A(\Delta t)^{-p}
\int_{-\infty}^{\infty} d(\Delta s)\,
\exp\!\left[iq\Delta s-\frac{\Delta s^2}{4A(\Delta t)}\right].
\]
The Gaussian integral is elementary:
\[
\int_{-\infty}^{\infty} dx\,
e^{iqx-x^2/(4A)}
=
\sqrt{4\pi A}\,e^{-q^2A}.
\]
Hence
\[
\widetilde C(q,\Delta t)
=
\widetilde C_0\,
A(\Delta t)^{-p+1/2}\,
e^{-q^2A(\Delta t)},
\]
where \(\widetilde C_0=\mathcal C_0\sqrt{4\pi}\). Therefore, in the diagonal approximation,
\begin{equation}
\label{eq:Gamma_n_appendix}
\left\langle \xi_n^{u,{\rm act}}(t)\xi_m^{u,{\rm act}}(t')\right\rangle
\simeq
\delta_{nm}\,\Gamma_n(\Delta t),
\end{equation}
where $\Gamma_n(\Delta t) = \Gamma_0\,q_n^2\,A(\Delta t)^{-p+1/2}\,e^{-q_n^2A(\Delta t)}$. Eq.~\eqref{eq:Gamma_n_appendix} is the projected microscopic memory kernel. It shows that the
active forcing becomes mode dependent in two ways: through the prefactor \(q_n^2\) associated with
the tangent projection, and through the time-dependent Gaussian cutoff
\(e^{-q_n^2A(\Delta t)}\).


The exact projected kernel \(\Gamma_n(\Delta t)\) is generally not a single exponential in time.
However, for the purpose of obtaining a compact analytical theory for the steady-state mode
variance, we replace it by the minimal single-timescale approximation
\[
\Gamma_n(\Delta t)
\;\longrightarrow\;
q_n^2 D_0\,e^{-q_n^2\sigma^2}\,e^{-|\Delta t|/\tau_{\rm eff}}.
\]
This replacement preserves the two most robust structural features of the microscopic projection:

\begin{itemize}
\item the tangent projection generates the overall factor \(q_n^2\),
\item the finite interaction range generates the Gaussian mode cutoff \(e^{-q_n^2\sigma^2}\).
\end{itemize}

The parameter \(\tau_{\rm eff}\) should therefore be interpreted as an effective memory time of the
\emph{projected} active forcing, rather than as a direct microscopic persistence time of the ABPs.
This leads to the effective closure used in the main text,
\[
\left\langle \xi_n^{u,{\rm act}}(t)\xi_m^{u,{\rm act}}(t')\right\rangle
=
\delta_{nm}\,
q_n^2 D_0 e^{-q_n^2\sigma^2}e^{-|\Delta t|/\tau_{\rm eff}}.
\]

\section{Evaluation of the mode variance}
\label{app:variance_derivation}

In this appendix we evaluate the mode variance corresponding to the linear mode equation
\[
\zeta\,\dot a_n(t)+\kappa q_n^4 a_n(t)=\xi_n^{\rm th}(t)+\xi_n^{\rm act}(t).
\]
Fourier transforming in time gives
\[
(\kappa q_n^4-i\omega\zeta)\,\tilde a_n(\omega)
=
\tilde \xi_n^{\rm th}(\omega)+\tilde \xi_n^{\rm act}(\omega),
\]
so that
\[
\tilde a_n(\omega)
=
H_n(\omega)\,
\big[\tilde \xi_n^{\rm th}(\omega)+\tilde \xi_n^{\rm act}(\omega)\big],
\]
where $H_n(\omega)={1}/({\kappa q_n^4-i\omega\zeta})$.
Hence
\[
|H_n(\omega)|^2
=
\frac{1}{\kappa^2 q_n^8+\zeta^2\omega^2}
=
\frac{1}{\zeta^2(\omega^2+\omega_n^2)},
\]
with $\omega_n = \kappa q_n^4/\zeta$.


The thermal tangent noise has spectrum
\[
S_n^{\rm th}(\omega)=2k_BT\,\zeta\,q_n^2.
\]
For the effective active kernel used in the main text,
\[
\left\langle \xi_n^{\rm act}(t)\xi_n^{\rm act}(0)\right\rangle
=
q_n^2D_0e^{-q_n^2\sigma^2}e^{-|t|/\tau_{\rm eff}},
\]
the Wiener--Khinchin theorem gives the active power spectrum
\[
S_n^{\rm act}(\omega)
=
q_n^2D_0e^{-q_n^2\sigma^2}
\frac{2\tau_{\rm eff}}{1+\omega^2\tau_{\rm eff}^2}.
\]


The steady-state variance is
\[
\langle |a_n|^2\rangle
=
\int_{-\infty}^{\infty}\frac{d\omega}{2\pi}\,
|H_n(\omega)|^2
\big[S_n^{\rm th}(\omega)+S_n^{\rm act}(\omega)\big]
\equiv I_n^{\rm th}+I_n^{\rm act}.
\]

The thermal contribution is
\[
I_n^{\rm th}
=
\int_{-\infty}^{\infty}\frac{d\omega}{2\pi}\,
\frac{2k_BT\,\zeta\,q_n^2}{\kappa^2q_n^8+\zeta^2\omega^2}.
\]
Using
\[
\int_{-\infty}^{\infty}\frac{d\omega}{\omega^2+\omega_n^2}
=
\frac{\pi}{\omega_n},
\]
one obtains
\[
I_n^{\rm th}
=
\frac{k_BT\,q_n^2}{\zeta\,\omega_n}
=
\frac{k_BT}{\kappa q_n^2}.
\]

The active contribution is
\[
I_n^{\rm act}
=
q_n^2D_0e^{-q_n^2\sigma^2}
\int_{-\infty}^{\infty}\frac{d\omega}{2\pi}\,
\frac{2\tau_{\rm eff}}
{(1+\omega^2\tau_{\rm eff}^2)(\kappa^2q_n^8+\zeta^2\omega^2)}.
\]
Write the denominator in terms of \(\omega_n=\kappa q_n^4/\zeta\):
\[
\kappa^2q_n^8+\zeta^2\omega^2
=
\zeta^2(\omega^2+\omega_n^2).
\]
Then
\[
I_n^{\rm act}
=
\frac{q_n^2D_0e^{-q_n^2\sigma^2}}{\zeta^2}
\int_{-\infty}^{\infty}\frac{d\omega}{2\pi}\,
\frac{2\tau_{\rm eff}}
{(1+\omega^2\tau_{\rm eff}^2)(\omega^2+\omega_n^2)}.
\]
The standard integral
\[
\int_{-\infty}^{\infty} d\omega\,
\frac{1}{(\omega^2+a^2)(\omega^2+b^2)}
=
\frac{\pi}{ab(a+b)}
\qquad (a,b>0)
\]
with \(a=1/\tau_{\rm eff}\) and \(b=\omega_n\) yields
\[
I_n^{\rm act}
=
\frac{q_n^2D_0e^{-q_n^2\sigma^2}}{\zeta^2}\,
\frac{\tau_{\rm eff}}{\omega_n(1+\tau_{\rm eff}\omega_n)}.
\]
Using \(\omega_n=\kappa q_n^4/\zeta\), this becomes
\[
I_n^{\rm act}
=
\frac{1}{\kappa q_n^2}\,
\frac{D_0\tau_{\rm eff}}{\zeta\,[1+\tau_{\rm eff}\omega_n]}\,
e^{-q_n^2\sigma^2}.
\]

Combining the thermal and active pieces gives
\begin{equation}
\label{eq:variance_appendix}
\langle |a_n|^2\rangle
=
\frac{1}{\kappa q_n^2}
\left[
k_BT
+
\frac{D_0\tau_{\rm eff}}{\zeta\,[1+\tau_{\rm eff}\omega_n]}
e^{-q_n^2\sigma^2}
\right].
\end{equation}
This is the expression quoted in the main text.


If a uniform backbone tension \(\tau\) is included, the mode relaxation rate is modified to
\[
\omega_n(\tau)=\frac{\kappa q_n^4+\tau q_n^2}{\zeta}.
\]
The thermal part of the variance is then
\[
\langle |a_n|^2\rangle_{\rm th}
=
\frac{k_BT}{\kappa q_n^2+\tau},
\]
and the active contribution is obtained by replacing \(\omega_n\to\omega_n(\tau)\) in the
frequency integral. The result is
\[
\langle |a_n|^2\rangle
=
\frac{1}{\kappa q_n^2+\tau}
\left[
k_BT
+
\frac{D_0\tau_{\rm eff}}{\zeta\,[1+\tau_{\rm eff}\omega_n(\tau)]}
e^{-q_n^2\sigma^2}
\right].
\]
This is the tension-generalized expression quoted in the main text.

\section{Radius of gyration from tangent-mode amplitudes}
\label{app:rg_from_tangent}
We work in the weak-bending regime, writing the polymer contour as
\begin{equation}
\mathbf{r}(s)=x(s)\,\hat{\mathbf{x}}+\mathbf{r}_\perp(s),
\end{equation}
with $s\in[0,L]$ and transverse tangent field $\mathbf{u}_\perp(s)\equiv \partial_s\mathbf{r}_\perp(s)$.
For free ends we expand $\mathbf{u}_\perp$ in the orthonormal cosine basis
\begin{equation}
\mathbf{u}_\perp(s)=\sum_{n\ge 1}\mathbf{a}_n\,\phi_n(s),\qquad
\phi_n(s)=\sqrt{\frac{2}{L}}\cos(q_n s),
\end{equation}
with $q_n=\frac{n\pi}{L}$, so that $\int_0^L ds\,\phi_n(s)\phi_m(s)=\delta_{nm}$.
Throughout we assume mode decoupling $\langle \mathbf{a}_n\cdot \mathbf{a}_m\rangle=\delta_{nm}\langle|\mathbf{a}_n|^2\rangle$.

\subsection{Transverse contribution $R_{g,\perp}^2$}
Integrating the tangent expansion gives
\begin{equation}
\mathbf{r}_\perp(s)=\int_0^s ds'\,\mathbf{u}_\perp(s')
=\sum_{n\ge 1}\mathbf{a}_n\,\sqrt{\frac{2}{L}}\frac{\sin(q_n s)}{q_n}.
\label{eq:rperp_app}
\end{equation}
The transverse radius of gyration is
\begin{equation}
R_{g,\perp}^2\equiv \frac{1}{L}\int_0^L ds\,\Big\langle\big|\mathbf{r}_\perp(s)-\mathbf{r}_{\perp,\rm cm}\big|^2\Big\rangle,
\end{equation}
where $\mathbf{r}_{\perp,\rm cm}\equiv \frac{1}{L}\int_0^L ds\,\mathbf{r}_\perp(s).$
Using Eq.~\eqref{eq:rperp_app}, mode decoupling, and $\frac{1}{L}\int_0^L ds\,\sin^2(q_n s)=\frac12$, we obtain 
\begin{equation}
\frac{1}{L}\int_0^L ds\,\langle|\mathbf{r}_\perp(s)|^2\rangle
=\sum_{n\ge 1}\frac{\langle|\mathbf{a}_n|^2\rangle}{Lq_n^2}.
\end{equation}
The CM term contributes only for odd $n$, since
\begin{equation}
\int_0^L ds\,\sin(q_n s)=\frac{1-(-1)^n}{q_n}
=\begin{cases}
0,& n\ \text{even},\\[4pt]
\frac{2}{q_n},& n\ \text{odd}.
\end{cases}
\end{equation}
A short calculation gives
\begin{equation}
\langle|\mathbf{r}_{\perp,\rm cm}|^2\rangle
=\sum_{n\ {\rm odd}}\frac{8}{L^2 q_n^4}\,\langle|\mathbf{a}_n|^2\rangle.
\end{equation}
Combining the two pieces yields the weighted sum rule
\begin{equation}
R_{g,\perp}^2
=\sum_{n\ge 1}\frac{\langle|\mathbf{a}_n|^2\rangle}{L\,q_n^{2}}\,W_n,
\qquad
W_n=
\begin{cases}
1, & n\ \text{even},\\[6pt]
1-\dfrac{8}{n^2\pi^2}, & n\ \text{odd},
\end{cases}
\label{eq:rg_perp_weighted_app}
\end{equation}
where we used $q_n=n\pi/L$ to rewrite the odd-mode correction.

\subsection{Longitudinal contribution $R_{g,\parallel}^2$}
Inextensibility implies $(\partial_s x)^2+|\mathbf{u}_\perp|^2=1$. Therefore,
\begin{equation}
\partial_s x=\sqrt{1-|\mathbf{u}_\perp|^2}\simeq 1-\frac12|\mathbf{u}_\perp|^2,
\end{equation}
which gives $x(s)=s-\epsilon(s)$ with 
\begin{equation}
\epsilon(s)\equiv \frac12\int_0^s ds'\,|\mathbf{u}_\perp(s')|^2.
\label{eq:eps_def_app}
\end{equation}
The longitudinal radius of gyration is
\begin{equation}
R_{g,\parallel}^2\equiv \frac{1}{L}\int_0^L ds\,\Big\langle\big(x(s)-x_{\rm cm}\big)^2\Big\rangle,
\end{equation}
where $x_{\rm cm}=\frac{1}{L}\int_0^L ds\,x(s)=\frac{L}{2}-\bar{\epsilon},$
with $\bar{\epsilon}\equiv \frac{1}{L}\int_0^L ds\,\epsilon(s)$.
Expanding to first order in $\epsilon$ (i.e.\ keeping the leading $\mathcal{O}(|\mathbf{u}_\perp|^2)$ correction) gives
\begin{equation}
\big(x(s)-x_{\rm cm}\big)^2
=\left(s-\frac{L}{2}\right)^2
-2\left(s-\frac{L}{2}\right)\big(\epsilon(s)-\bar{\epsilon}\big)
+\mathcal{O}(\epsilon^2).
\end{equation}
Upon integration over $s$, the $\bar{\epsilon}$ term drops out because $\int_0^L ds\,(s-L/2)=0$, so
\begin{equation}
R_{g,\parallel}^2
\simeq {\frac{1}{L}\int_0^L ds\left(s-\frac{L}{2}\right)^2}
-\frac{2}{L}\int_0^L ds\left(s-\frac{L}{2}\right)\langle\epsilon(s)\rangle.
\label{eq:rgpara_lin_app}
\end{equation}
To evaluate $\langle\epsilon(s)\rangle$, note that mode decoupling implies
\begin{equation}
\langle|\mathbf{u}_\perp(s)|^2\rangle
=\frac{1}{L}\sum_{n\ge 1}\langle|\mathbf{a}_n|^2\rangle\Big[1+\cos(2q_n s)\Big],
\label{eq:uperp2_profile_app}
\end{equation}
since $\phi_n(s)^2=(2/L)\cos^2(q_ns)=(1/L)[1+\cos(2q_ns)]$.
Using Eq.~\eqref{eq:eps_def_app} then gives
\begin{equation}
\langle\epsilon(s)\rangle
=\sum_{n\ge 1}\langle|\mathbf{a}_n|^2\rangle
\left[
\frac{s}{2L}+\frac{\sin(2q_n s)}{4Lq_n}
\right].
\label{eq:eps_mean_app}
\end{equation}
Substituting Eq.~\eqref{eq:eps_mean_app} into Eq.~\eqref{eq:rgpara_lin_app} and using $q_n=n\pi/L$ yields
\begin{equation}
R_{g,\parallel}^2
\simeq \frac{L^2}{12}
-\frac{L}{12}\sum_{n\ge 1}\langle|\mathbf{a}_n|^2\rangle
+\frac{L}{4\pi^2}\sum_{n\ge 1}\frac{\langle|\mathbf{a}_n|^2\rangle}{n^2}.
\label{eq:rg_parallel_app}
\end{equation}
Equation~\eqref{eq:rg_parallel_app} captures the leading reduction of the projected size.
Corrections of order $\epsilon^2$ correspond to fluctuations of the contraction field and are quartic in mode amplitudes.

\subsection{Total $R_g^2$}
Within this approximation the total squared radius of gyration decomposes as
\begin{equation}
R_g^2 \simeq R_{g,\parallel}^2 + R_{g,\perp}^2,
\end{equation}
with $R_{g,\perp}^2$ from Eq.~\eqref{eq:rg_perp_weighted_app} and $R_{g,\parallel}^2$ from Eq.~\eqref{eq:rg_parallel_app}.
In practical evaluations, the mode sums are truncated at the microscopic cutoff set by discretization.

\section{Exact finite-$N$ reconstruction of the radius of gyration}
\label{app:finiteN_rg}

In this appendix we derive an exact finite-$N$ representation of the radius of gyration for the
discrete bead--spring chain used in the simulations. The purpose of this construction is to provide
a benchmark reconstruction of $R_g$ that does not rely on the continuum weak-bending approximation.

Consider a chain of $N+1$ monomers at positions $\{\mathbf r_i\}_{i=0}^{N}$ and define the bond
vectors
\[
\Delta \mathbf r_k \equiv \mathbf r_{k+1}-\mathbf r_k,
\qquad k=0,\dots,N-1.
\]
The radius of gyration can be written exactly as
\begin{equation}
\label{eq:rg_discrete_def_app}
R_g^2
=
\frac{1}{(N+1)^2}\sum_{i=0}^{N}\sum_{j=i+1}^{N}
\left\langle |\mathbf r_j-\mathbf r_i|^2 \right\rangle .
\end{equation}
Since
\[
\mathbf r_j-\mathbf r_i=\sum_{k=i}^{j-1}\Delta \mathbf r_k,
\]
Eq.~\eqref{eq:rg_discrete_def_app} depends only on the bond vectors.

We expand the bond vectors in a discrete cosine basis,
\begin{equation}
\label{eq:bond_mode_expansion_app}
\Delta \mathbf r_k
=
\frac{1}{2}\mathbf b_0
+\sum_{n=1}^{N-1}\mathbf b_n
\cos\!\left[\frac{n\pi}{N}\left(k+\frac12\right)\right].
\end{equation}
Summing Eq.~\eqref{eq:bond_mode_expansion_app} from $k=i$ to $j-1$ yields
\begin{equation}
\label{eq:rji_discrete_app}
\mathbf r_j-\mathbf r_i
=
\frac12 \mathbf b_0 (j-i)
+\sum_{n=1}^{N-1}\mathbf b_n\,
\frac{\sin(n\pi j/N)-\sin(n\pi i/N)}
{2\sin(n\pi/2N)}.
\end{equation}
Equation~\eqref{eq:rji_discrete_app} is exact for the finite chain and makes explicit the geometric
difference between the discrete reconstruction and its continuum large-$N$ approximation.

Substituting Eq.~\eqref{eq:rji_discrete_app} into Eq.~\eqref{eq:rg_discrete_def_app} and carrying
out the sums over monomer indices gives a quadratic form in the bond-mode covariance matrix,
\begin{multline}
\label{eq:rg_discrete_matrix_app}
R_g^2
=
W_{00}\,\langle \mathbf b_0\!\cdot\!\mathbf b_0\rangle
+\sum_{n\ge 1}W_{nn}\,\langle \mathbf b_n\!\cdot\!\mathbf b_n\rangle\\
+\sum_{n\ge 1}W_{0n}\,\langle \mathbf b_0\!\cdot\!\mathbf b_n\rangle
+\sum_{n<m}W_{nm}\,\langle \mathbf b_n\!\cdot\!\mathbf b_m\rangle.
\end{multline}
The coefficients \(W_{nm}\) are purely geometric finite-$N$ weights determined by the discrete
basis and the definition of $R_g$. In practice, they may be evaluated exactly from the discrete sums
appearing in Eqs.~\eqref{eq:rg_discrete_def_app} and \eqref{eq:rji_discrete_app}, yeilding:

\begin{align}
{W_{00} }&{= \frac{N(N+2)}{48}, }\\
{W_{nn} }&{= \frac{N(N+1) - 2 \mathcal{O}_n \cot^2\left(\frac{n\pi}{2N}\right)}{8(N+1)^2\sin^2\left(\frac{n\pi}{2N}\right)}, }\\
{W_{0n} }&{= 
\begin{cases}
-\frac{N\cos\left(\frac{n\pi}{2N}\right)}{4(N+1)\sin^2\left(\frac{n\pi}{2N}\right)} & n \text{ even},\\[10pt]
0 & n \text{ odd},
\end{cases} }\\
{W_{nm} }&= 
\begin{cases}
-\frac{\cos\left(\frac{n\pi}{2N}\right)\cos\left(\frac{m\pi}{2N}\right)}{2(N+1)^2\sin^2\left(\frac{n\pi}{2N}\right)\sin^2\left(\frac{m\pi}{2N}\right)} & n, m \text{ odd},\\[10pt]
0 & \text{otherwise},
\end{cases}
\end{align}

where $n<m$ and the indicator function $\mathcal{O}_n$ is $1$ for odd $n$ and $0$ for even $n$. Thus, the non-diagonal couplings link the zeroth mode exclusively to even modes, and odd modes exclusively to each other.
When the full simulation-measured covariance matrix is inserted into Eq.~\eqref{eq:rg_discrete_matrix_app}, the result reproduces the directly measured $R_g$ essentially exactly.

Equation~\eqref{eq:rg_discrete_matrix_app} shows that the continuum weak-bending reconstruction of
Sec.~III.E omits two distinct ingredients: finite-chain geometric weights and off-diagonal
cross-correlations between discrete bond modes. The exact finite-$N$ reconstruction therefore serves as a benchmark for diagnosing whether discrepancies in $R_g$ arise from the continuum approximation, from neglected cross-correlations, or from genuinely missing physics such as activity-induced bond stretching.

\end{document}